\documentclass[12pt]{article}
\usepackage{amsmath}
\usepackage{graphicx}
\usepackage{longtable}
\usepackage{amssymb}

\begin{document}

\footnotesize{\noindent {\it Scientific Inquiry}  {\bf 7}, No. 1, June 30, pp. 25 - 50 (2006).  \  \   \   \   \  \   \  \  \   IIGSS Academic Publisher}

\begin{center}
\section*{\LARGE {Reasons for nuclear forces in light of  \\ the
constitution of the real space} }
\end{center}

\medskip

\begin{center}
{\bf Volodymyr Krasnoholovets}
\end{center}

\begin{center}
{Institute for Basic Research \\ 90 East Winds Court, Palm Harbor,
FL 34683, U.S.A.}
\end{center}

\begin{flushright}
    {22 April 2005 \ \ }
\end{flushright}

\begin{center}
\textbf{Abstract}
\end{center}

\hspace*{\parindent} The concept of microstructure of the real
space, considered as a mathematical lattice of cells (the
tessellattice), and notions of canonical particles and fields,
which are generated by the space, are analyzed. Submicroscopic
mechanics based on this concept is discussed and employed
for in-depth study of the nucleon-nucleon interaction. It is
argued that a deformation coat is developed in the real space
around the nucleon (as is the case with any other canonical
particle such as electron, muon, etc.) and that they are the deformation
coats that are responsible for the appearance of nuclear forces.
One more source of nuclear forces is associated with inerton
clouds, excitations of the space tessellattice (the excitations
are a substructure of nucleons' matter waves), which accompany
moving nucleons as in the case of any other canonical particles.
Two nuclear systems are under consideration: the deuteron and a
weight nucleus. It is shown that a weight nucleus is a cluster of
interacting protons and neutrons. The condition of the cluster
stability is obtained in the framework of the statistical
mechanical approach. On the whole, the paper proposes a radically
new approach to a very old unsolved problem, the origin of the
nuclear forces, allowing the understanding mechanism(s) occurring
at low energy nuclear transmutations.

\bigskip

\textbf{Key words:} inertons, nucleons, nuclear forces, quantum
mechanics, space structure, tessellattice

\medskip

\textbf{PACS:} 21.30.-x nuclear forces; \ 21.80.+a hypernuclei; \
21.60.Gx cluster models; \  24.80.+y nuclear tests of fundamental
interactions and symmetries;  \ 24.60.-k statistical theory and
fluctuations; \ 11.90.+t other topics in general theory of fields
and particles

\newpage

\begin{flushright} \footnotesize{O Vasi\d{s}\d{t}has, your greatness
is spread like sun's light,   \\        \ \ \ \ \ \ \ \ \ \quad
\quad \quad \quad \quad \quad\quad \quad \quad \quad \quad\quad is
deep like ocean, has speed like air.  {\kern 1pt}\quad \quad \ \ \
\ \ \ \ \ \ \ \ \ \ \ \ \\ \hspace{5cm} \quad \ $The$ {\it ${\d
R}gveda$}, \ 7.33.8 \quad\quad}
\end{flushright}

\section*{1. Introduction and preliminaries}

\subsection*{1.1. Conventional views}
\hspace*{\parindent} Although nuclear physics is a well-developed
branch of modern physical science, avenues for the resolution of
the problem of the origin of nuclear forces are still beyond
understanding. At present the model approach to deriving nuclear
forces from the quark-quark interaction prevails among
researchers. Nevertheless, such an approach is open to question,
especially owing to the confinement problem, which is the most
difficult one for quantum chromodynamics (QCD).

It is a matter of fact that the understanding how QCD works
remains one of the great puzzles of many-body physics. Indeed, the
degrees of freedom observed in low energy phenomenology are
totally different from those appearing in the QCD Lagrangian. In
the case of many-nucleon systems, the question of the origin of
the nuclear energy scale is immediately arouse: the typical energy
scale of QCD is on the order of 1 GeV, though the nuclear binding
energy per particle is very small, on the order of 10 MeV. Is
there some deeper insight from which this scale naturally arises?
Or the reason should one searches in complicated details of near
cancellations of strongly attractive and repulsive terms in the
nuclear interaction?

This large separation between the hadronic energy scale and the
nuclear binding scale has led to an alternative approach,
especially at the description of the physics of small {\it A}
nuclei. Namely, effective field theory techniques, which arise from
chiral symmetry, allow quantitative calculations of energy
parameters of such nuclei. In particular, these methods have been
employed to account for the physics of pions in the context of
chiral perturbation theory. Currently efforts are made to combine
these effective field theories with the low energy constants from
QCD, which then might be considered as first principles
calculations. A further fundamental progress is expected from the
lattice field theory and the use of super computers.

Nevertheless, recent high precision measurements (Gilman and Gross
[1]) of the deuteron electromagnetic structure functions ({\it A},
{\it B} and $T_{20}$) extracted from high-energy elastic {\it ed}
scattering, and the cross sections and asymmetries extracted from
high-energy photodisintegration $\gamma + d \rightarrow  n + p$,
have been reviewed and compared with the theory. The theoretical
consideration included nonrelativistic and relativistic models
using the traditional meson and baryon degrees of freedom,
effective field theories, and models based on the underlying quark
and gluon degrees of freedom of QCD, including nonperturbative
quark cluster models and perturbative QCD. The conclusion has been
drawn that analyzes of the elastic {\it ed} scattering experiment
and the photodisintegration experiment require very different
theoretical approaches.

In other words, the rigorous results of work [1] demonstrate that
QCD and the meson theory seem disagree. Hence the origin of
nuclear forces in a unperturbed nucleus is still unclear.

\subsection*{1.2. Hadronic mechanics}

\hspace*{\parindent} Although QCD and quantum field theories are
rather considered as independent disciplines, they, however, are
strongly based on nonrelativistic and relativistic quantum
mechanics. Santilli proposed (see, e.g. Ref. [2,3]) the broadening
foundations of quantum mechanics known under the title of {\it
hadronic mechanics}. This mechanics has been developing now by
numerous researchers both theoreticians and experimentalists. In
particular, hadronic mechanics has successfully been applied for
the study of stimulated nuclear transmutations in the range of low
and intermediate energies [2].

Santilli has shown [2] how conventional quantum mechanics is
insufficient to solve its most fundamental problem: the physical
origin of the strong interaction of nuclear constituents (a number
of potentials used does not give any explanation whatever of the
origin of attraction between nucleons); the theoretical prediction
of quantum mechanics for the correct representation of the
deuteron in its ground state misses 2.3\% of the experimental
value. There are also other strong insufficiencies of quantum
mechanics for the representation of nuclear data [2].

Regarding a possibility of deriving nuclear forces from the
quark-quark interaction, Santilli [2] reasonably remarks that
quarks can only be defined in a mathematical unitary space that
has no direct connection to an actual physical reality. Then he
continues: ``It is known by experts that, because of the
impossibility of being defined in our space-time, \textit{quarks
cannot have any scientifically meaningful gravitation}, and their
``masses" are pure mathematical parameters in the mathematical
space of SU(3) with no known connection to our space-time."
Indeed, quarks cannot be defined through special relativity and
its fundamental Poincar${\rm \acute{e}}$ symmetry. Therefore, this
means that mass cannot be introduced as the second order Casimir
invariant, however, this is only the necessity for mass to exist
in our space-time.

In other words, the basic challenge of deducing the gravitational
interaction from quarks seems completely unfeasible.

Taking into account the mentioned fundamental difficulties,
Santilli, developing hadronic mechanics, started from the
hypothesis that in a system of strongly interacting particles the
total Hamiltonian cannot be subdivided to kinetic and potential
parts. Hadronic mechanics is constructed via a nonunitary
transform of orthodox quantum mechanics, namely, the unitary
character appears only at a distance larger than the radius of
nuclear forces $r_{\rm n.{\kern 1pt}f.} \approx 10^{-15}$ m:
\begin{displaymath}
U \times U^{\dagger}
 = {\hat I} \left\{
\begin{array}{ll}
    \neq I &  \textrm{for} \quad  r \ll r_{\rm n.{\kern 1pt}f},   \\
       = I &  \textrm{for} \quad  r \gg r_{\rm n.{\kern 1pt}f.}.  \\
\end{array} \right.
\end{displaymath}

Then a simple yet in fact effective, realization of this
assumption for the relativistic treatment of a system of two
nucleons, i.e. deuteron, is Animalu-Santilli isounit [2]
\begin{equation*}
{\hat I}_{12} =\prod\limits_{k} {\rm Diag} \left( n^2_{1{\kern
1pt}k}, n^2_{2{\kern 1pt}k},n^2_{3{\kern 1pt}k},n^2_{4{\kern
1pt}k}\right) \times \exp \left( N \int d {\kern 1pt}
v {\kern 2pt} {\hat \psi}^{\dag}_{\uparrow k} ({\bf{r}}) \times
{\hat \psi}_{\downarrow {\kern 0.5 pt}k} ({\bf{r}})\right)
\end{equation*}
where $k=1,{\kern 1pt}2$. The functions $n^2_{ik}$ represent the
shape of the nucleon and characterizes semiaxes of spheroidal
ellipsoids normalized to the values for the perfect shape and to
the volume preserving simple conditions.

Such a transform allows the introduction of a modernized Lie
product, Pauli matrices, Dirac equation, etc. So, the conventional
Hamiltonian of relativistic quantum mechanics and the appropriate
spinors are transformed to a new Santilli's isomathematical
presentaions. The mathematics developed enabled the calculation of
basic parameters of nuclear systems, which exactly agreed the
experimental data [2].

Unlike conventional quantum mechanics that operate with point
particles and their appropriate wave packets, or wave functions,
hadronic mechanics deals with the extended particles that feature
peculiar shapes in our space-time.

Hadronic mechanics indeed yields remarkable predictions both in
chemical and nuclear physics. In particular, it has predicted new
stable chemical species such as hydrogen and etc., which have been
verified experimentally, and which are specified by the stronger
binding energy of electrons with atoms (around 2 times larger than
in case of conventional species). A similar hypothesis has been
offered for the structure of the neutron: the neutron has been
treated as a strongly coupled proton-electron pair. This enabled
one to associate the nuclear forces with the pure Coulomb
interaction between protons and neutrons. Thus in hadronic
mechanics the origin of nuclear forces is complete plain: this is
the usual Coulomb interaction between nucleons.

\subsection*{1.3. Submicroscopic consideration}

\hspace*{\parindent} QCD, effective quantum theories and hadronic
mechanics widely employ notions and means of orthodox quantum
mechanics, first of all Schr\"odinger's wave $\psi $-function and
Dirac's spinor formalisms, which themselves are exposed to
significant conceptual difficulties [4].

Nonlocality, which the wave $\psi $-function introduced, and
action at-a-distance forces are the most challenging questions of
orthodox quantum mechanics [4]. In fact, in quantum mechanics all
potentials are treated as static and long-distance: a parabolic
potential in the harmonic oscillator problem, the Coulomb
potential in the hydrogen atom problem, etc. (see also Arunasalam
[5]).

So, all modern quantum theories including quantum mechanics, QCD,
hadronic mechanics and etc. imply long-range action. Therefore, in
this respect they do not differ from the phenomenological Newton's
gravitational law, which indeed establishes the link $G{\kern 1pt}
M_{1} M_{2} /r$ between two distant objects, but does not account
for the mechanism that realizes the interaction. General
relativity also does not account for the origin of Newton's
potential, but simply employs it as a starting point complicating
the theory.

Moreover, no one of quantum theories available does pay any
attention to the background of systems studied, i.e. the structure
and peculiarities of the real physical space. The theories are
developed in abstract spaces: energy, momentum, phase, Hilbert and
so on. Instead of the background space they use such complete
undetermined notions as a ``physical vacuum" or/and an ``aether"
providing them with every possible and imaginary properties. It
seems the aforementioned Santilli's quotation regarding quarks as
objects that are not determined in the space-time is the apt turn
of phrase, which emphasizes the validity of our criticism.

Having overcome difficulties of quantum physics caused by the
deficiency of knowledge about the background, Bounias and the
author [6-9] have recently undertaken an in-depth study of space
as it is. Starting from topology, set theory and fractal geometry
we have revised the principal mathematical notions, by which time
the mathematical and physical literature has presented.

We  have based our consideration [6-9] on the assumption that the
real space, i.e. a 3D space or a 4D space-time, is not a dim
vacuum but a quantum substrate that shares discrete and continual
properties. Similarly to condensed matter physics in which the
availability of a regular/irregular atom lattice plays the
fundamental role, we have introduced a special mathematical
lattice of topological balls, which characterizes the real space
in detail. This space generates matter (particles) and provides
for the creation of physics laws. In particular, the interaction
of a particle with the surrounding space, cells of the
mathematical lattice called the ``tessellattice",  has to produce
short-range action, i.e. excitations of the tessellattice, which
are capable of carrying the interaction for long distances from
the particle, as in the case of excitations of the crystal lattice
of a solid.

A particle appears as a local deformation of the tessellattice.
The deformed cell being stable represents an actual core of any
canonical particle.

It is assumed [10] that the size of a superparticle, the building
block of the tessellattice, is on the order of $10^{-30}$ m. In
fact, it is known that on this scale three of four physical
interactions (electromagnetic, weak and strong, except for the
gravitational interaction) should come together. Besides, in
particle physics researchers use the notion of an abstract
superparticle whose different states are electron, positron, muon,
quarks, etc.

A cell deformation of the tessellattice means the induction of
mass in the appropriate cell and the value of mass is directly
proportional to the degree of the cell deformation. Such
deformation can only be the fractal deformation [7,8]. Besides,
the class of leptons (electron, muon and $\tau$-lepton) is
characterized by the fractal reduction of the initial volume of a
cell, though the class of quarks features the fractal expansion of
the initial volume of a cell [8,9].

The electric charge is associated with a quantum of fractality
that is complete located in one cell [8,9]. The charge state is
located on the surface of the particled cell. A positive charged
particle has naturally been determined as an object whose surface
covered by protuberances, then a negative charged particle is
specified by the surface covered by cavities [8,11]. A detailed
theory of the charge and the submicroscopic interpretation of the
Maxwell equations have been developed in paper [11].

The theory of the real space allows the construction of a
mechanics of particles, which incorporates the interaction of a
moving particle with the tessellattice essentially representing
the degenerate, or non-manifest, state of the space. In other
words, this is a dynamics of the mathematical lattice, the
tessellattice, which being developed will complete substitute all
modern theories and their attempts aimed at the construction of a
unified theory of the nature.

\medskip

In the present work we briefly describe submicroscopic quantum
mechanics developed in the real space, which has been constructed
by the author [12-16]. It should be emphasized that the theory has
successfully been verified experimentally [17-19] and agree well
with many data. Then the results obtained in the framework of
submicroscopic approach are applied to the consideration of
nuclei. That is, we will study peculiarities of the nucleon
structure, which generate nuclear forces, and disclose the inner
reasons for these forces. After that we will investigate actual
trajectories of nucleons in the deuteron, i.e. trajectories of the
deuteron's nucleons in the real space. Furthermore, we will
analyze the origins of major potentials, which are present in
weight nuclei and which show that a nucleus should be treated as a
cluster of nucleons.

\section*{2. Dynamics of tessellattice}

\hspace*{\parindent} We have shown [6-9] that the universe indeed
can be constructed of nothing, or in other words, the
universe allows the description in the form of a mathematical
lattice of empty sets, i.e. degenerate cells that can be called
superparticles. Obviously, a dynamics of the mathematical lattice,
called the tessellattice [6-9], should include all possible kinds
of transformations and movements allowable by the combinations of
mathematical rules in the structure of packed cells taking into
account possible cells' shapes and kinds of symmetries, the
inversion of cell deformation (e.g., the transition quark -
lepton), the exchange of fractal deformations, and the real motion
of deformed cells, i.e. particles, in the tessellattice. The
theory of dynamics of the tessellattice is still in the
rudimentary state and its further development will need the
tremendous work of numerous researchers.

At the same time, the motion of a particle in the tessellattice
should be related to pure mechanics, but what kind of a mechanics?
The problem has been studied by the author in a series of works
[4,10-19]. The research has shown that the mechanics of particles
in the tessellattice (i.e. in the real space) constitutes
\textit{submicroscopic deterministic quantum mechanics}, because
it is easily reduced to the formalism of conventional quantum
mechanics: Schr{\"o}dinger's [12,13] and Dirac's [14], which is
developed in abstract phase or Hilbert spaces.

We recall that quantum mechanics, as such, being based on
indeterminism is constructed in the phase space and establishes
probabilistic links between measurable characteristics of the
system studied. The most essential characteristics, or parameters,
of particles are their energy, momentum, moment of momentum, and
position. Submicroscopic deterministic quantum mechanics, which is
called submicroscopic mechanics below, makes broader the
framework of orthodox quantum mechanics. The deterministic
approach has been constructed in the real space in which the
particle's parameters mentioned above are determined. Thus
submicroscopic mechanics being developed in the real space allows
causal links between the given particle parameters.

In submicroscopic mechanics, the creation of a canonical particle
is treated as the appearance of a local deformation in the
tessellattice, i.e. a stable change in the volume of a
superparticle is associated with the formation of a massive
particle. The mass is determined as the ratio between the initial
volume of a superparticle in the degenerate state of the
tessellattice and the volume of the same superparticle in the
deformed state, $M_{{\kern 0.4pt}0} \propto {\kern 1pt} {\kern
1pt} {\kern 1pt} {\kern 1pt} {\kern 1pt} {\mathcal{V}}^{\left(
{\rm sup} \right)}/{\kern 1pt} {\kern 1pt} {\mathcal{V}}^{\left(
{\rm part} \right)}$. A quasi-particle, or elementary excitation
of the space tessellattice, is also associated with the appearance
of a local deformation that is small and unstable in comparison
with a particle; the mass of an excitation of the tessellattice
$m_{{\kern 1pt}0} \propto {\kern 1pt} {\kern 1pt} {\kern 1pt}
{\kern 1pt} {\kern 1pt} {\mathcal{V}}^{\left( {\rm sup}
\right)}/{\kern 1pt} {\kern 1pt} {\mathcal{V}}^{\left( {\rm excit}
\right)}$. Therefore, $M_{{\kern 0.4pt} 0} \gg m_{{\kern 0.3pt}0}
$.

Note that in the case of mechanics, there is no need to take into
account peculiarities of the volume decrease of a cell, which
results in the induction of mass in the corresponding cell.
Although the pure volume decrease is not sufficient for providing
a cell with mass; only the fractal-related decrease of the volume
of the cell causes an actual deformation that is associated with
mass [7,9].

In solid state physics, the occurrence of a foreign particle in
the crystal lattice automatically induces a range of a deformation
in the crystal lattice, i.e. the so-called deformation coat, which
covers from several lattice constants to several tens of the
lattice constants. By analogy, we should construct a deformation
coat around a canonical particle in the tessellattice, which plays
the role of a screen that separates the particle from the
degenerate space. Indeed a particled cell should be characterized
by a strong local deformation. Hence the tessellattice should
responses to this deformation in such a way that all cells in a
range $R$ around the particled cell become deformed down to the
boundary where edge cells are bounded by a rupture of the
remaining fractality. We equal the size $R$ of the deformation
coat to the Compton wavelength $\lambda _{{\kern 1pt} \rm Com} =
h/M_{0} {\kern 1pt} c$ of the particle in question [14]. Hence the
radius of the deformation coat coincides with the value of
$\lambda _{{\kern 1pt} \rm Com} $. Since cells, or superparticles,
inside the coat are deformed, they possess mass as well as the
particle that is found in the center of the formation.

When a particle starts to move with a velocity $v $, it pushes its
way through coming fluctuating superparticles and thus the
particle collides with superparticles. Owing to the interaction
with coming superparticles the particle has to emit elementary
excitations. As the excitations represent inert properties of the
particle, they were called ``inertons" [12-14]; inertons are
generated due to the resistance of the space that the moving
particle experiences (see also Ref. [7]). The moving particle also
pulls its deformation coat, but superparticles in the coat remain
motionless. The superparticles' massive state travels by a relay
mechanism, i.e. the particle adjusts surrounding superparticles to
the deformation coat state; it is suggested that this state is
adjusted with a speed no less than the speed of light \textit{c}.

The particle emits inertons along a section $\lambda /2$ of its
path and hence it loses the velocity, $v \to 0$; on the next
section $\lambda /2$ the elastic tessellattice sends inertons
backward to the particle and the particle velocity is restored to
the value $v $ and so on. The section $\lambda $ is the equivalent
of the de Broglie wavelength of the particle. In such a manner
inertons make up a substructure of the matter waves and therefore
they should be considered as carriers of the quantum mechanical
force, or quantum mechanical potential, generated by the particle
in the ambient space. The velocity of inertons cannot be lesser
than the speed of light \textit{c}. Thus inertons allow us to
completely remove long-range action from quantum mechanics.

Inertons transmitting the energy and the momentum also carry
fragments of local deformation of the tessellattice, i.e. they
atomize the particle's inert mass in the surrounding of the
particle. This means that the mentioned quantum mechanical
potential (or the deformation of the space) should be identified
with the gravitational potential of the particle. This problem has
been studied in detail in papers [21,16] where the process of inerton
emission and the reentry of inertons into the particle have also
been investigated. The transition to the formalism of orthodox
quantum mechanics takes place by relationships
\begin{equation}
\label{eq1} E = h{\kern 1pt} {\kern 1pt} \nu ,\quad \quad \lambda
= \frac{{h}}{{p}},
\end{equation}

\noindent which are derived in the frame of the submicroscopic
consideration.

Relationships (\ref{eq1}) were first written down for a particle
by Louis de Broglie in 1924 (see, e.g. Ref. [20]) and then were
derived by the author [12,13] in the framework of the
submicroscopic approach briefly stated herein. The availability of
relationships (\ref{eq1}) enables [20] a simple derivation of the
Schr\"odinger wave equation.

\begin{figure}
\begin{center}
\includegraphics[scale=0.8]{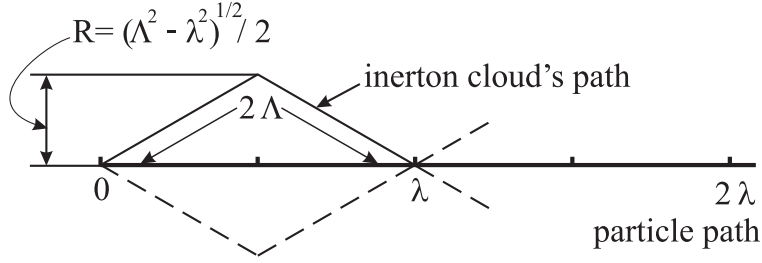}
\caption{\small{Schematic representation of the section $\lambda $
along the particle path and the effective path length $2\Lambda $
of the particle's inerton cloud.}} \label{Figure 1}
\end{center}
\end{figure}

In the submicroscopic approach the parameter \textit{E} is equal
to the total kinetic energy of the particle $\tfrac{{1}}{{2}}M_{0}
v ^{2}/\sqrt {1 - v ^{2}/c^{2}} ; \quad \nu $ denotes the
frequency of oscillations of the particle along its path, namely,
the frequency of the oscillation of the particle's velocity due to
the periodic emission and reabsorption of inertons, $v \to 0 \to v
\to 0 \to ...$ (besides, $\nu $ is connected with the period of
collisions $T$ of the particle with its inerton cloud, $\nu =
1/2T$); the de Broglie wavelength $\lambda $ represents here the
spatial period, or amplitude, of oscillations of the particle
along its path; $p = M_{\kern 1pt 0} v / \sqrt {1 - v ^{2}/c^{2}}$
is the total momentum of the particle.

The mentioned parameter \textit{T} also enables other
presentations: $T = v \lambda$ and $T = c\Lambda$ where $\Lambda $
is the free path length of the cloud of inertons, which the cloud
passes between two sequential collisions with the particle
($\Lambda $ can also be called the amplitude of the inerton
cloud). Similarly, $\lambda $ can be called the free path length
of the particle. Such a presentation makes it possible to consider
the particle's behavior in terms of kinetic theory. The particle's
path and the particle's parameters $\lambda $ and $\Lambda $ are
sketched in Figure 1. As follows from Figure 1, the range covered
by the inerton cloud around the particle can be estimated by a
radius $R = \sqrt {\left( {\Lambda /2} \right)^{2} - \left(
{\lambda /2} \right)^{2}} $, i.e. $R \sim \Lambda $.

So, the availability of the inerton cloud enclosing a canonical
particle endows the abstract (probabilistic) wave $\psi $-function
with an actual physical sense: The $\psi $-function covers the
range of the space disturbed by the moving particle, $R \sim
\Lambda $, and sets up links between the parameters of the
particle and those of the particle's inerton cloud. In other
words, the $\psi$-wave function describes the inerton field of the
particle, which spreads out of the particle up to the range $ \sim
\Lambda $.

Equating relations $T = v \lambda $ and $T = c\Lambda $, we get
the relationship
\begin{equation}
\label{eq2} \Lambda = \lambda {\kern 1pt} {\kern 1pt}
\frac{{c}}{{v} }{\kern 4pt}
\end{equation}
that relates the particle parameters $\lambda $ and $v $ to
corresponding parameters of the particle inerton cloud, $\Lambda $
and $c$. Relationship (\ref{eq2}) can be rewritten by using the
Compton wavelength $\lambda _{{\kern 1pt} {\kern 1pt} \rm Com} $
of the particle,
\begin{equation}
\label{eq3} \Lambda = \lambda _{{\kern 1pt} {\kern 1pt} \rm Com}
{\kern 2pt}{\kern 1pt}\frac{{c^{2}}}{{v ^{2}}}{\kern 4pt}.
\end{equation}

We see from expression (\ref{eq3}) that when the particle velocity
$v $ tends to $c$, the inerton cloud becomes practically closed in
a range embraced by the Compton wavelength, or in other words, the
deformation coat (or the space crystallite) that is developed
around the particle in the space tessellattice.

The submicroscopic quantum theory stated above could be usefully
employed in the studies of those quantum systems to which
conventional quantum mechanics is also applied. Perhaps the main
value of the submicroscopic approach is its short-range action,
which so far has been overlooked in the other concepts. Besides,
submicroscopic mechanics could solve such difficult theoretical
problems as disagreement on the Schr\"odinger equation and Lorentz
invariance [13], the interpretation of the spin [14,8], the nature
of the phase transition that occurs in the quantum system under
consideration when we pass from the description based on the
Schr\"odinger formalism to that resting on the Dirac one [14], the
origin of gravity [16,21], the nature of the photon [22,11] and so
on. Moreover, the theory has predicted the existence of a new
physical field -- the inerton field -- that then has successfully
been fixed experimentally [17] (see also Refs. [18,19]).

All these results allow us to anticipate that the submicroscopic
approach should also be useful when considering the intricate
challenge associated with the nature of nuclear forces, the more
so, as experimentalists who investigate low energy nuclear
reactions claim that they fix a new ``strange" radiation at the
transmutation of nuclei [23,24]. Clearly, that ``strange"
radiation was nothing else as inertons, which radiated from the
appropriate inerton clouds of nucleons when the latter rearranged
in nuclei.

\section*{3. Deformation coat of the nucleon}

\hspace*{\parindent} Since superparticles in the deformation coat
surrounding a particle possesses masses, they should fall within
collective vibrations by analogy with massive entities (atoms,
molecules) in the crystal lattice. This analogy has also allowed
us to call the deformation coat the space crystallite and then to
investigate its vibratory mode [14]. In the crystallite,
vibrations of all superparticles co-operate and the total energy
of superparticles, which is equal to the total energy of the
particle, $M_{{\kern 1pt}0} {\kern 1pt} c^{2}$, is quantized,
\begin{equation}
\label{eq4} \hbar {\kern 1pt}{\kern 1pt} \omega _{k_{0}}  =
M_{{\kern 1pt}0} {\kern 1pt} c^{2}
\end{equation}
where $k_{{\kern 1pt} 0} = 2\pi /\lambda _{{\kern 1pt} \rm Com} $
is the wave number and $\omega _{k_{{\kern 1pt} 0}}  = ck_{{\kern
1pt} 0} $ is the cyclic frequency of an oscillator in the
$k$-space (the quantity $\lambda _{{\kern 1pt} \rm Com} $ is the
amplitude of the oscillator, which is given by the crystallite
size, i.e. the Compton wavelength). For the moving particle
expression (\ref{eq4}) is transformed to
\begin{equation}
\label{eq5} \hbar {\kern 1pt}{\kern 1pt} \omega _{k_{v} }  =
M{\kern 1pt} c^{2}
\end{equation}
where $M = M_{{\kern 1pt}0} /\sqrt {1 - v ^{2}/c^{2}} $ and
$k_{{\kern 1pt}v} = 2\pi /\left( {\lambda _{{\kern 1pt}\rm Com}
/\sqrt {1 - v ^{2}/c^{2}}} \right)$. The crystallite travels
together with the particle, namely, coming superparticles adjust
to the massive state in the range covered by the crystallite size
$\lambda _{{\kern 1pt}\rm Com} / \sqrt {1 - v ^2/c^2}$ by a relay
mechanism with a speed no less than the speed of light $c$.

Figure 2 depicts the deformation coat, or crystallite, that is
progressing around the canonical particle in the degenerate
tessellattice. Figure 2 rather sketches the spatial pattern of a
lepton (electron, positron, muon, etc.).

\begin{figure}
\begin{center}
\includegraphics[scale=0.8]{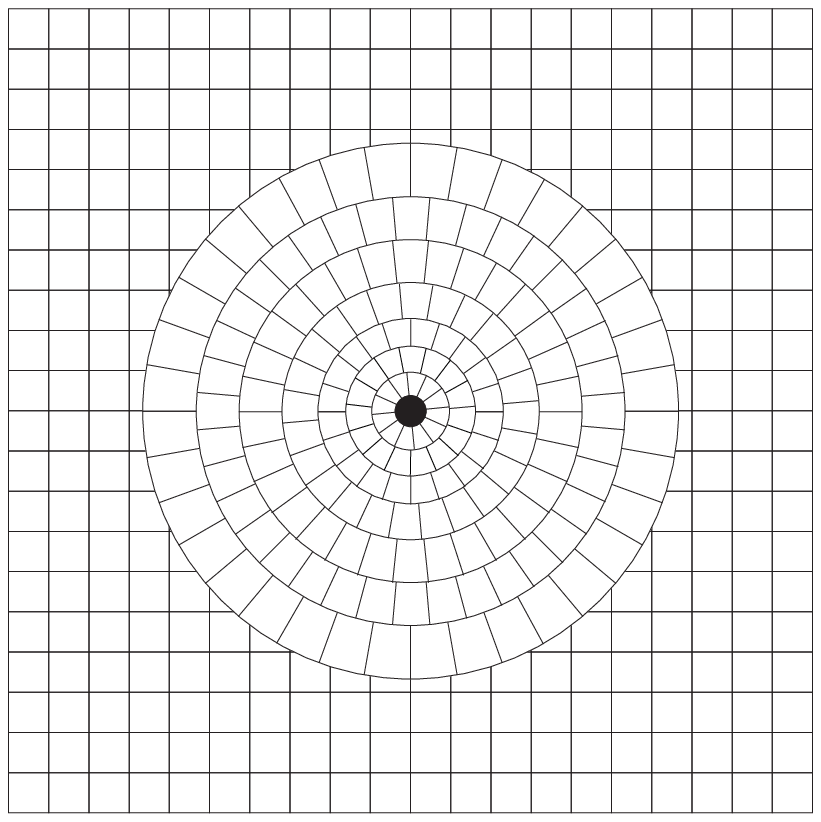}
\caption{\small{Canonical particle surrounded by its deformation
coat, or space crystallite, in the space tessellattice.}}
\label{Figure 2}
\end{center}
\end{figure}

In the case of a quark we should anticipate a similar picture,
though the quark itself and superparticles in its deformation coat
should be characterized by a deformation that is inverse in
comparison with that that specifies leptons and superparticles in
the lepton's coat. This is the only feature that distinguishes a
family of quarks from that of leptons and only this feature can
account for the non-stability of isolated quarks. The only way for
quarks to survive is their integration into groups of a smaller
total size, hadrons, which have to lower the degree of the inverse
deformation of space associated with the quark and thus should
stabilize them. Here we will not analyze peculiarities of the
spatial pattern of quarks in a hadron and the motion of quarks in
it (though the motion should not in principle be disobedient to
the rules of submicroscopic mechanics stated above). We shall
restrict our consideration to the fact that the hadron features
first quarks' generalized deformation coat and then an outer, or
usual deformation coat shown in Figure 3 in which superparticles
of the space tessellattice have a smaller size than that in the
degenerate state.

For the nucleon we may ascribe the radius $R_{{\kern 1pt }\rm c}
\simeq 0.4 \times 10^{ - 15}$ m to the quark's generalized
deformation coat, because on this size, as is known (see, e.g.
Ref. [25]), nuclear forces of attraction are converted into forces
of repulsion. The masses of nucleons, $M_{0{\kern 1pt}\rm neutron}
\cong M_{0{\kern 1pt} \rm proton} = M_{{\kern 1pt}0} = 1.67 \times
10^{ - 27}$ kg and therefore the nucleon's Compton wavelength
$\lambda _{{\kern 1pt} {\kern 1pt} \rm Com} = h/M_{0} {\kern 1pt}
c = 1.32 \times 10^{ - 15}$ m. This magnitude should be identified
with an actual radius of the (outer) deformation coat of the
nucleon. Figure 3 depicts the nucleon in the real space: there is
the main body (the hadron core) with the radius $R_{\rm c} \simeq
0.4 \times 10^{ - 15}$ m in the center part, which includes three
quarks, and the deformation coat (crystallite) surrounding the
core with the radius $\lambda _{{\kern 1pt} {\kern 1pt} \rm Com} =
1.32 \times 10^{ - 15}$ m.

\begin{figure}
\begin{center}
\includegraphics[scale=0.7]{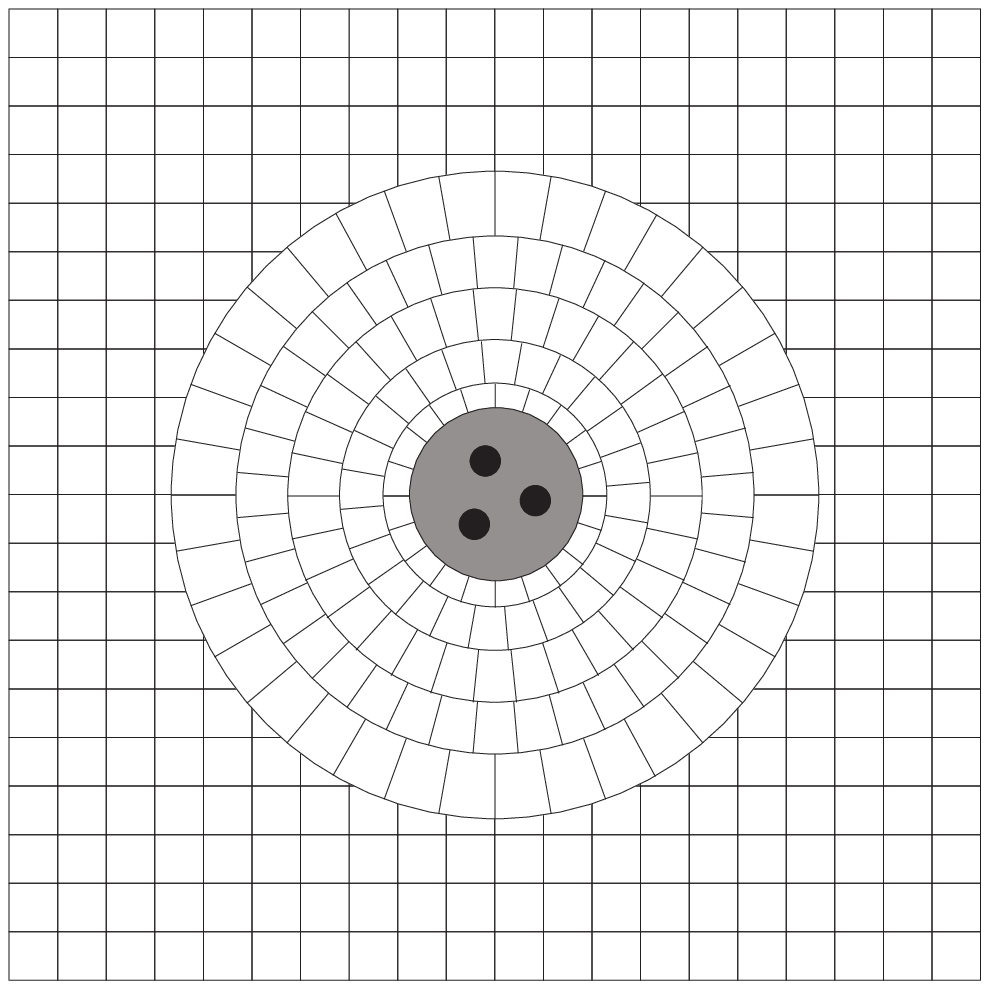}
\caption{\small{ Degenerate space and the nucleon, which consists
of the main body, or core (shaded) and the deformation coat that
surrounds the core.}} \label{Figure 3}
\end{center}
\end{figure}

Since the size of a nucleus $R \simeq R_{{\kern 1pt} 0} {\kern
1pt} A^{1/3}$ where $R_{{\kern 1pt} 0} = \left( {1.2\,\,\,{\rm
to}\,\,\,1.3} \right) \times 10^{ - 15}$ m and $A$ is the number
of nucleons, we may identify the parameter $R_{{\kern 1pt} 0} $
with the said Compton wavelength $\lambda _{{\kern 1pt} {\kern
1pt} \rm Com} $ and therefore $R_{{\kern 1pt} 0} $ becomes the
radius of the deformation coat of the nucleon. Beyond the coat
superparticles are found in the degenerate state where they do not
possess any mass.

\subsection*{\textbf{\textit{3.1. Behavior of the deformation coat in
the phase space}}}

\hspace*{\parindent} The value of the potential energy accumulated
in the deformation coat of a nucleon ($M_{0{\kern 1pt} \rm n}
{\kern 1pt} c^{2}$ for neutron and $M_{0{\kern 1pt} \rm p} {\kern
1pt} c^{2}$ for proton) stores up in the vibrating mode of the
coat, which shows expression (\ref{eq5}) that has been derived at
the consideration in the energy space (see Appendix in Ref. [14]).
Vibrations of atoms in the crystal lattice simplify periodical
oscillations of atoms around their equilibrium positions. In the
case of the deformation coat the vibratory process can occur
according to a similar scenario. Besides, owing to the spherical
form of the crystallite we may reduce the problem to the
well-known task of mathematical physics on small vibrations of a
gas contained in a sphere (see, e.g. Ref. [26]).

The velocities potential $u$ of the gas fulfills the wave equation
\begin{equation}
\label{eq6} \frac{{\partial ^{{\kern 1pt}2}u}}{{\partial {\kern
1pt} r^{2}}} + \frac{{2}}{{r}}\frac{{\partial {\kern 1pt} {\kern
1pt} u}}{{\partial {\kern 1pt} r}} =
\frac{{1}}{{c^{2}}}\frac{{\partial {\kern 1pt} {\kern 1pt}
u^{2}}}{{\partial {\kern 1pt} t^{2}}};
\end{equation}

\noindent here $u$ for radial vibrations is a function of $r{\kern
1pt}$ and $t$ alone where $r{\kern 1pt} $ is the distance from a
vibrating particle of the gas to the center of the sphere and $t$
is the time. Equation (\ref{eq6}) is solved by substitution
\begin{equation}
\label{eq7} u (r,{\kern 2pt} t)= w \left( {r} \right) {\kern 1pt}
\theta \left( {t} \right)  .
\end{equation}

The surface of the sphere is treated as a hard envelope and
therefore the normal component of the velocity is equal to zero,
which leads to the boundary condition
\begin{equation}
\label{eq8}
\left. {\frac{{\partial {\kern 1pt} {\kern 1pt} w\left( {r}
\right)}}{{\partial {\kern 1pt} {\kern 1pt} r}}} \right|_{\,\,r = R_{{\kern
1pt} 0}}  \, = 0
\end{equation}

\noindent
where $R_{{\kern 1pt} 0} $ is the radius of the spherical envelope (the
radius of the deformation coat). The solution for $w\left( {r} \right)$ is
\begin{equation}
\label{eq9} w\left( {r} \right) = C{\kern 1pt} {\kern 1pt} {\kern
1pt} \frac{{\sin\left( {\eta {\kern 1pt} {\kern 1pt} r + \varphi}
\right)}}{{r}},
\end{equation}

\noindent
or in the explicit form
\begin{equation}
\label{eq10} w_{s}  \left( {r} \right) = \frac{{1}}{{r}}{\kern
1pt} {\kern 1pt} {\kern 1pt} {\kern 1pt} {\kern 1pt} {\kern 1pt}
\sin\left( {\frac{{\mu _{{\kern 1pt} s}  + \alpha} }{{R_{{\kern
1pt} 0}} }{\kern 1pt} {\kern 1pt} {\kern 1pt} r} \right)
\end{equation}

\noindent
where
\begin{equation}
\label{eq11} \mu _{{\kern 1pt} s}  = \left( {s + \tfrac{{1}}{{2}}}
\right){\kern 1pt} {\kern 1pt} \pi - \varepsilon _{s}  ,
\end{equation}

\noindent
 $\alpha $ is the positive constant and the function $\varepsilon _{s}$
satisfies equation
\begin{equation}
\label{eq12} \varepsilon _{{\kern 1pt} s}  = \frac{{2}}{{\left(
{2s + 1} \right){\kern 1pt} {\kern 1pt} \pi} } + \frac{{4{\kern
1pt}\varepsilon _{{\kern 1pt} s} ^{2}} }{{3{\kern 1pt} {\kern 1pt}
\left( {2s + 1} \right){\kern 1pt} {\kern 1pt} \pi} }.
\end{equation}

The partial solution to the task is
\begin{equation}
\label{eq13} u_{{\kern 1pt} s} \left( {r,\,\,t} \right) = U_{s}
{\kern 1pt}{\kern 1pt}\frac{{1}}{{r}}{\kern 1pt} {\kern 1pt}
{\kern 1pt} {\kern 1pt} \sin\left( {\frac{{\mu _{{\kern 1pt} s}  +
\alpha} }{{R_{{\kern 1pt} 0}} }{\kern 1pt} {\kern 1pt} {\kern 1pt}
r} \right)\,\,\sin\left( {c{\kern 1pt}\frac{{{\kern 1pt} {\kern
1pt} \mu _{{\kern 1pt} s}  + {\kern 1pt} {\kern 1pt} \alpha
}}{{R_{{\kern 1pt} 0}} }{\kern 1pt} {\kern 1pt} {\kern 1pt} t +
\beta _{s} } \right).
\end{equation}

Although the total solution is $u = \sum\nolimits_{s} {u_{{\kern
1pt} s} }  $, we restrict our study by expression (\ref{eq13}) for
$u_{{\kern 1pt} s}  \left( {r,\,\,t} \right)$ as we assumed that
massive superparticles should vibrate in the deformation coat only
in one collective mode [14].

Nuclear forces appear when deformation coats of separate nucleons
overlap. Indeed, vibrating superparticles of one coat begin to
interact with those of the other coat. It is a matter of fact that
the interaction between two oscillators reduces the total energy
of the oscillators. In our situation the overlapping means that
boundary condition (\ref{eq8}) is destroyed: the derivative
$\partial {\kern 1pt} w/\partial {\kern 1pt} {\kern 1pt} r$
becomes other than zero at $r = R_{{\kern 1pt} 0} $. The
nucleon-nucleon interaction expends the potential $w\left( {r}
\right)$ of one participant beyond its spherical envelope inside
of the potential induced by the other participant. Therefore we
can expect that condition (\ref{eq8}) will now be realized at the
other effective distance $r = R_{{\kern 1pt} 0} + \Delta R$, i.e.
\begin{equation}
\label{eq14}
\left. {\frac{{\partial {\kern 1pt} {\kern 1pt} w\left( {r}
\right)}}{{\partial {\kern 1pt} {\kern 1pt} r}}} \right|_{\,\,r = R_{{\kern
1pt} 0} + \Delta R} \, = 0.
\end{equation}

Boundary condition (\ref{eq14}) alters arguments $\left( {\mu
_{{\kern 1pt} s}  + \alpha}  \right) r/R_{{\kern 1pt} 0} {\kern
1pt}$ and $ \left( {\mu _{{\kern 1pt} s} + \alpha} \right) c{\kern
1pt}t/R_{{\kern 1pt} 0} {\kern 1pt} {\kern 1pt} {\kern 1pt}$ in
expression (\ref{eq13}) to $\left( {\mu _{{\kern 1pt} s} + \alpha}
\right) r/ \left( {R_{{\kern 1pt} 0} + \Delta R} \right)$ and $c
\left( {\mu _{{\kern 1pt} s} + \alpha} \right) t/ \left(
{R_{{\kern 1pt} 0} + \Delta R} \right)$, respectively. This
enables us to rewrite the solution (\ref{eq13}) in the form
\begin{equation}
\label{eq15} u_{{\kern 0.5pt} s}  \left( {r,\,\,t} \right) = U_{s}
{\kern 1pt}{\kern 1pt} \frac{{1}}{{r}}{\kern 1pt} {\kern 1pt}
{\kern 1pt} {\kern 1pt} {\kern 1pt} \sin\left( {\frac{{\mu
_{{\kern 1pt} s}  + \alpha} }{{R_{{\kern 1pt} 0} + \Delta
R}}{\kern 1pt} {\kern 1pt} {\kern 1pt} r} \right)\,\, \sin\left(
{c\frac{{{\kern 1pt} {\kern 1pt} \mu _{{\kern 1pt} s}  + {\kern
1pt} {\kern 1pt} \alpha} }{{R_{{\kern 1pt} 0} + \Delta R}}{\kern
1pt} {\kern 1pt} {\kern 1pt} t + \beta _{s} } \right).
\end{equation}

Solution (\ref{eq15}) shows that the union of nucleons lowers
their energy. In fact the multipliers at time $t$ in expressions
(\ref{eq13}) and (\ref{eq15}) are the respective cyclic
frequencies $\omega _{{\kern 1pt} s}  $ and $\tilde {\omega
}_{{\kern 1pt} s}  $, that is,
\begin{equation}
\label{eq16} \omega _{{\kern 1pt} s}  = c {\kern 1pt} {\kern
1pt}\frac{{{\kern 1pt} {\kern 1pt} \mu _{{\kern 1pt} s}  + {\kern
1pt} {\kern 1pt} \alpha} }{{R_{{\kern 1pt} 0} }},\quad \quad
\tilde {\omega} _{{\kern 1pt} s}  = \omega _{{\kern 1pt} s}
\times \left( {1 - \frac{{\Delta R}}{{R_{{\kern 1pt} {\kern 1pt}
0} }}} \right).
\end{equation}

It is reasonable to assume that proper harmonic vibrations of
massive superparticles in the deformation coat occur at the
fundamental tone that is characterized by the frequency $\omega
_{{\kern 0.3pt} 1} =  {\kern 1pt} \left( {\mu _{{\kern 0.3pt} 1} +
\alpha}  \right) c{\kern 1pt} / R_{{\kern 1pt} 0} $ where $\mu
_{{\kern 0.3pt} 1} \simeq 4.494$ [27]. If we set $\alpha = 1.785$,
we indeed reach the equality
\begin{equation}
\label{eq17} \hbar {\kern 1pt}{\kern 1pt} \omega _{{\kern 0.4pt}
1} = M_{\kern 0.4pt 0{\kern 1pt}\rm  n} {\kern 1pt} c^{2} \cong
M_{\kern 0.4pt 0{\kern 1pt} \rm p} {\kern 1pt}{\kern 1pt} c^{2} =
938.26 {\kern 4pt}{\rm MeV}.
\end{equation}

Since the interaction of nucleons can reduce their energy by $W =
47$ MeV (in agreement with the Fermi gas model, see e.g. Ref.
[28]), we can write the equality
\begin{equation}
\label{eq18} \hbar {\kern 1pt}{\kern 1pt} \omega _{{\kern 0.3pt}
1} {\kern 1pt} {\kern 1pt} {\kern 1pt} \frac{{1}}{{1 + \Delta
R/R_{0}} } \simeq \left( {938.26 - 47} \right) {\rm MeV}
\end{equation}

\noindent that makes it possible to estimate an effective range of
the overlapping of deformation coats of two nucleons, $\Delta R
\simeq 0.053{\kern 1pt} R_{{\kern 1pt} 0} $. Such an overlapping
can virtually draw two nucleons together, but only a little. A
deeper penetration into the core size $R_{{\kern 1pt} c} $ can be
achieved only in the case of weight nuclei when the collective
motion of a great number of nucleons is allowed for.

Thus the study conducted in the framework of the phase space has
shown that the coupling of nucleons is a beneficial process. Now
let us take a look at the behavior of nucleons in the real space.

\subsection*{\textbf{\textit{3.2. Behavior of the deformation coat
in the real space}}}

\hspace*{\parindent} Submicroscopic mechanics briefly discussed
above has been constructed [10,12-15] as the kinetic theory of a
particle, which does not pay any attention to the reasons that
ensure the re-absorption of inertons emitted by the particle at
its motion.

The problem of reversion of inertons has been studied in Refs.
[16,21]. During the period of oscillations of a
particle/superparticle the value of its mass periodically changes
between $M_{\kern 0.5pt 0} /\sqrt {1 - v ^{2}/c^{2}} $ and
$M_{\kern 0.5pt 0} $, i.e., any motion of a particle is associated
with the periodical disintegration of its inert mass. Since by
definition the mass characterizes a local deformation of the space
tessellattice, i.e. $M_{{\kern 0.4pt} 0}$ reflects a change in the
volume of a degenerate superparticle, the mass oscillation implies
periodical changes of the particle/superparticle's volume between
the oblate state, ${\mathcal{V}}/\sqrt {1 - v ^{2}/c^{2}}$, and
the rest one, ${\mathcal{V}}$. The potential part of the total
energy of the oscillating superparticle corresponds to the
deviation of the superparticle from its equilibrium position in
the space tessellattice.

In the case of the nucleon's crystallite, we can follow a similar
pattern: Vibrations of superparticles in the crystallite means
that the values of their masses oscillate in such a way that the
local deformation (the mass) periodically passes into the local
deformation of the tessellattice as the whole (called the rugosity
in Refs. [16,21]). In other words, the contraction of one
superparticle (the mass) competes with the shift of the
superparticle from its equilibrium position (the rugosity) in the
tessellattice. So, we can construct the following specific
Lagrangian that describes the behavior of superparticles in the
deformation coat (compare with Ref. [16])
\begin{equation}
\label{eq19} L = \sum\limits_{\vec n} {\left(
{\tfrac{{1}}{{2}}{\kern 1pt}{\kern 1pt}\dot {m}_{{\kern 1pt} {\bf
n}}^{2} + \tfrac{1}{2}{\kern 1pt}{\kern 1pt} {\dot {\bf \Xi}
_{{\kern 1pt}{\bf n}}}^{\kern 1pt 2} - c{\kern 1pt} {\kern 1pt}
\dot {m}_{{\kern 1pt} {\bf n}}  \nabla {\kern 1pt}{{\bf \Xi}}
_{{\kern 1.5pt} {\bf n}}} \right)}.
\end{equation}
Here $m_{{\kern 1pt} {\bf n}} $ is the mass of a superparticle
that is specified by the radius vector ${\bf n}$ and ${{\bf \Xi}}
_{{\kern 1pt} {\bf n}} $ is the deviation of the said
superparticle from its equilibrium position in the deformation
coat.

Euler-Lagrange equations of motion written for $m_{{\kern 1pt} n}
$ and ${{\bf \Xi}} _{{\kern 1.5pt}{\bf n}}$ are reduced to two
uncoupled wave equations below where we omit the index ${\bf n}$
(see technical details in Refs. [16,21])
\begin{equation}
\label{eq20} \frac{{\partial ^{{\kern 1pt} 2}m}}{{\partial {\kern
1pt} r^{2}}} + \frac{{2}}{{r}}\frac{{\partial {\kern 1pt} {\kern
1pt} m}}{{\partial {\kern 1pt} r}} =
\frac{1}{c^2}\frac{\partial^{\kern 1pt 2} {\kern 1pt} m}{\partial
{\kern 1pt} t^2},
\end{equation}
\begin{equation}
\label{eq21} \frac{{\partial ^{{\kern 1pt}2}{\kern 1pt}{\Xi}}
}{{\partial {\kern 1pt} r^{{\kern 1pt}{\kern 1pt}2}}} +
\frac{{2}}{{r}}\frac{{\partial {\kern 1pt} {\kern 1pt} {\Xi}}
}{{\partial {\kern 1pt} r}} =
\frac{{1}}{{c^{2}}}\frac{\partial^{\kern 1pt 2} {\kern 1pt} \Xi
}{{\partial {\kern 1pt} t^2}}.
\end{equation}

In equations (\ref{eq20}) and (\ref{eq21}) owing to the spherical
symmetry $m$ and $\Xi$ are functions of the distance $r$ alone
($r$ is the distance from the superparticle under consideration to
the center of the sphere) and time $t$. Once again, the local
radial translational deformation $\Xi$, which is the only
component of the vector ${\bf \Xi}$,  is treated as a collective
parameter of the vibrating deformation coat describing a deformed
range of the space as a whole; that is why the term "rugosity" of
the space has been used to name it [16,21].

Equations (\ref{eq20}) and (\ref{eq21}) are similar in form with
Eq. (\ref{eq6}), however, they do not require such a hard boundary
condition as expression (\ref{eq8}) prescribes. Taking into
account that values $m$ and $\Xi $ should oscillate in opposite
phases, we arrive at appropriate boundary condition
\begin{equation}
\label{eq22} \left. {m\left( {r,\,\,t} \right)} \right|_{{\kern
1pt} r = R_{{\kern 1pt} 0}}  = 0;\quad \quad \left. {\Xi \left(
{r,\,\,t} \right)} \right|_{{\kern 1pt} r = 0} = 0.
\end{equation}

Then the solution for $m$ becomes
\begin{equation}
\label{eq23} m{\kern 1pt} {\kern 1pt} \left( {r,\,\,t} \right) =
m_0\frac{r_{01}}{r}\left|{\kern 1pt}\cos {\kern 1pt} \frac{{\pi
{\kern 1pt} r}}{{2R_{{\kern 1pt} 0}} }\right|  \left| {{\kern 1pt}
\cos {\kern 1pt} \frac{{\pi {\kern 1pt} t}}{{2 T }}{\kern 1pt}
{\kern 1pt}} \right|;
\end{equation}
\begin{equation}
\label{eq24} \Xi{\kern 1pt} {\kern 1pt} \left( {r,\,\,t} \right) =
\Xi_{\kern 0.6pt 0} {\kern 1pt} {\frac {r_{02}}{r}} {\kern 2pt}
\left| {\kern 1pt} \sin {\kern 1pt} \frac{{\pi {\kern 1pt}
r}}{{2R_{{\kern 1pt} 0}} }{\kern 1pt} {\kern 1pt} \right|  \left|
{\kern 1pt} \sin {\kern 1pt} \frac{{\pi {\kern 1pt} {\kern 1pt}
t}}{2T}{\kern 1pt} {\kern 1pt} \right|;
\end{equation}

\noindent  here $r\leq R_0$; the parameters $r_{01}$ and
$r_{0{\kern 0.2pt}2}$ are normalized constants; $m_{{\kern
0.4pt}0}$ is the maximum value of mass, which is realized at the
center of the sphere; $\Xi_{{\kern 0.5pt}0}$ is the maximum value
of the rugosity, which is reached at the boundary $r= R_{{\kern
0.5pt} 0}$.

Solutions (\ref{eq23}) and (\ref{eq24}) bring out the behavior of
superparticles in the deformation coat in the real space. In fact,
the most interesting for us the solution (23) shows that the mass
$m$ of a superparticle located at a distance $r$ from the center
periodically oscillates with $t$, i.e. the volume of the
superparticle oscillates. From the viewpoint of the constitution
of the space this means that one kind of the deformation is
periodically transformed into the other one, namely, a local
deformation of the tessellattice is transferred to a local
translational radial deformation and on the contrary. In the
deformation coat the mass distribution is given by the respective
amplitude in expression (\ref{eq23}),
\begin{equation}
\label{eq25} \mathcal{A}_{{\kern 1pt} \rm mass} \propto
\frac{m_{{\kern 0.5pt} 0}}{r}{\kern 1pt}{\kern 1pt}\cos{\kern 1pt}
\frac{{\pi {\kern 1pt} r}}{{2R_{{\kern 1pt} 0}} }{\kern 1pt}
{\kern 1pt}.
\end{equation}

In the state of the pure translational deformation the
superparticle under consideration does not possess any mass,
$m\left( {r,{\kern 1pt} {\kern 1pt} {\kern 1pt} {\kern 1pt} t}
\right) = 0$, i.e. its volume equals the volume of a superparticle
located in the degenerate tessellattice, but it is slightly
shifted from the equilibrium position. Therefore, the deformation
coat that consists of massive superparticles represents an actual
deformed range of the space surrounding the nucleon. The degree of
the spatial deformation is distributed in the coat in line with
the amplitude $\mathcal{A}_{{\kern 1pt} \rm mass} $ (\ref{eq25}).

In such a manner the deformation coat with the range of $R_{{\kern
1pt} 0} = \lambda _{{\kern 1pt} \rm Com} = h/M_{0} {\kern 1pt} c$
represents the actual deformation potential of the nucleon. This
is the mass field, or deformation field, which is responsible for
the availability of attractive nuclear forces in nuclei. The
nucleon-nucleon interaction changes the boundary conditions
(\ref{eq22}); in particular, for the mass the condition is
affected to
\begin{equation}
\label{eq26} \left. {m\left( {r,\,\,{\kern 1pt} t} \right)}
\right|_{{\kern 1pt} r = R_{{\kern 1pt} 0} + \Delta R} = 0,
\end{equation}
that is, the overlapping of two coats extends the range of action
of the massive (deformation) potential. This brings about the
replacement of $R_{{\kern 1pt} 0} $ for $R_{{\kern 1pt} 0} +
\Delta R$ in the solution (\ref{eq23}).

The amplitude of inerton cloud $\Lambda$ (3) much exceeds the
object size. Because of that, the interaction between objects is
carried out by their inerton clouds, which overlap similarly to
the overlapping of deformation coats of interacting nucleons.
Consequently, the nature of macroscopic gravity and that of
microscopic nuclear forces are completely the same, they vary only
in scale. In fact, in the case of the gravitation [16,21], we have
obtained solutions for the inerton cloud, which have the form of
expressions (23) and (24); the difference is only in scale: If the
boundary of the deformation coat is found at the distance of $R_0
= \lambda_{\rm Com}$ from the center of a nucleon, the boundary
$\Lambda$ of the nucleon's inerton cloud is determined by
relationship (3). The interaction of overlapping inerton clouds of
two distant particles with masses $M_{1} $ and $M_{2} $ is
characterized by the energy $G M_{1} M_{2} /r$ that is extremely
small in comparison with the interaction of overlapping
deformation coats of nucleons in a nucleus (several tens of MeV).

Thus the phenomenon of the attraction, in essence, is caused by
the contraction of the space that surrounds physical objects. In
both the deformation coat and the inerton cloud, masses represent
deformed cells (or superparticles) of the space tessellattice and
hence these domains of the space being in the contracted state
induce actual deformation potentials of the objects.

The overlapping of the deformation coats means that one core comes
under the influence of the deformation coat surrounding the other
core. In other words, due to the overlapping superparticles in the
deformation coats become more deformed, or massive, and the value
of the appropriate deformation should depend on the degree of the
overlapping of the coats, which is proportional to the radius
change $\Delta R$ evaluated above. Using the experimental data on
the minimum value of the potential well ($W = $ 35 to 47 MeV), we
can evaluate an increase in the total mass of superparticles found
in the deformation coat. Since the total energy $M_{0} {\kern 1pt}
c^{2} \to M_{0} {\kern 1pt} c^{2} - \Delta E$, we may write for
the total mass of superparticles in the coat: $M_{0} \to M_{0} +
\Delta M$ where $\Delta M = M_{{\kern 1pt}0} \times \Delta
E/\left( {M_{0} {\kern 1pt} c^{2}} \right) \simeq \left(
{0.037\;\,{\rm to} \;\,0.05} \right) \times M_{0}$. Hence the
notion of a "potential well" implies that in the range of space
covered by the well, spatial blocks, i.e. superparticles, are
found in a more contracted state than in the space beyond the
potential well.

\section*{4. Nucleons in deuteron}

\hspace*{\parindent} Although the deuteron is the simplest nuclear
system, the thorough theory of the deuteron is still far being
complete. Of course, many parameters of the deuteron have already
been clarified; for instance, the low boundary of the minimum of
the potential well, $W_{0} = 35$ MeV, the binding energy, $\Delta
E = 2.22$ MeV, the effective radius in triplet and singlet states,
$R_{{\kern 1pt} \rm d {\kern 2pt} t ( s )}$, etc. (see, e.g. Ref.
[29]). Nevertheless, the inner reasons for the proton-neutron
coupling and many peculiar details associated with the behavior of
nucleons in the deuteron are not yet entirely known.

Let us now look at the problem of the deuteron from the viewpoint
of the constitution of the real space that is developing. It is
reasonable to associate the potential well with the overlapping of
deformation coats of two nucleons. Since the radius of the nucleon
coat is $R_{{\kern 1pt} 0} = h/M_{\rm n{\kern 1pt} (p)} {\kern
1pt} c = 1.32 \times 10^{ - 15}$ m, the inequality $R_{\rm {\kern
0.5pt} d} > R_{{\kern 0.5pt} 0}$, where $R_{\rm {\kern 0.5pt} d}$
stands for the radius of the deuteron, should mean that the
deuteron's proton and neutron are found outside the overlapping of
their deformation coats for a long period of time. Such a pattern
implies that the space tessellattice must take part in the
oscillating process of the nucleons. The deuteron's proton and
neutron feature virtually the same mass and velocity, therefore,
their interaction should be like the behavior of elastic balls in
a viscous substrate [21]. In other words, the elastic space plays
the role of a spring that holds each of the two nucleons in place.

Let us study the behavior of these two nucleons in term of the
submicroscopic approach. Any particle, including a nucleon, moves
emitting and reabsorbing its inerton cloud. These processes occur
in the section that equals the de Broglie wavelength $\lambda$.
Emitted inertons are elastically reflected from the space
tessellattice at a distance of $\Lambda \sim \lambda {\kern 1pt}
{\kern 1pt} c/v $ (\ref{eq2}) from the particle and then come back
to the particle. Note that at the range of $\Lambda$, local
deformations of the space, which are carried by inertons, are
converted into the rugosity of the space, which is a kind of
a tension of the space, i.e. roughly speaking,
$m \to \Xi$. The space tessellattice straightens the rugosity and
initiates the reverse inerton motion: the rugosity passes into the
local deformation, $\Xi \to m$, and so on.

Let us now calculate the de Broglie wavelength in a nucleus. The
specific energy per nucleon $E_{{\kern 1pt} 1}$ in a nucleus  is
around 8 MeV [30] and therefore from equality
\begin{equation}
\label{eq27} \frac{{M_{{\kern 1pt} 0{\kern 1pt}}  {\kern 1pt} v
^{2}}}{{2\sqrt {1 - v ^{2}/c^{2}}} } = E_{{\kern 1pt} 1}
\end{equation}

\noindent we get for the nucleon velocity $v \approx \sqrt {2E_{
1} /M_{{\kern 1pt}0}}  \approx 5.8 \times 10^{7}$ m/s. Hence the
de Broglie wavelength of the nucleon $\lambda = h/Mv \approx 0.6
\times 10^{ - 14}$ m. These parameters allow us to calculate the
amplitude of the inerton cloud of the nucleon by relation
(\ref{eq2}), $\Lambda = \lambda {\kern 2pt} {\kern 1pt} c/v
\approx 3.5 \times 10^{ - 14}$ m.

For the deuteron, putting the binding energy $\Delta E = 2.22$
MeV, which  is equal to the kinetic energy of a nucleon $E_{{\kern
1pt} 1} $, we get from expression (\ref{eq27}): $v \simeq 2.06
\times 10^{7}$ m/s, $\lambda \simeq 1.79 \times 10^{ - 14}$ m, and
$\Lambda \simeq 2.6 \times 10^{ - 13}$ m.

It is interesting to note that the estimated values of $R_{{\kern
1pt} 0}$ and $\Lambda $ imply that in an atom whose size is $\sim
10^{ - 10}$ m neither deformation coats of the nucleus' nucleons
nor inerton clouds of nucleons can reach electron orbits. This
fact on its own means the nucleus does not hold the electrons in
the orbital position. At the same time, for an electron in an atom
the Compton wavelength (i.e. the electron's deformation coat)
$\lambda _{{\kern 1pt}\rm Com}^{( {\rm electr})} = h/M_{\rm
electr} {\kern 1pt} c \simeq 2.4 \times 10^{ - 12}$ m, the de
Broglie wavelength $\lambda _{{\kern 1pt} \rm electr} = h/M_{\rm
electr}{\kern 1pt} v \simeq 3.5 \times 10^{ - 10}$ m, and the
range covered by the electron's inerton cloud $R_{{\kern 1pt} \rm
electr} \approx \tfrac 12 \Lambda _{\rm electr} = \lambda _{{\kern
1pt} \rm electr} {\kern 1pt}{\kern 1pt} c/2{\kern 1pt}v \sim
10^{-8}$ m. So the amplitude of the inerton cloud of an atom's
electron far exceeds the atom size. As has recently been shown
[22,11], photons represent nothing else but electromagnetically
polarized inertons. That is why in the atom, only
electromagnetically polarized inerton clouds of electrons capture
the atomic nucleus. In other words, polarized inertons of atomic
electrons directly interact with the nucleus. Hence contrary to
the consensus, those are electrons that generate the Newton (and
Coulomb) potential $1/r$ of an atom.

The difficult problem of the motion of nucleons in the deuteron,
which takes into account their interaction with the space, can be
reduced to a more simple study that employs the kinetic approach
first proposed in submicroscopic mechanics of a free particle
[12-14]. Let us construct the Lagrangian that makes allowance for
the actual contraction of a moving object, i.e. the nucleus in our
case (the so-called relativistic Lagrangian), the overlapping of
deformation coats of nucleons (i.e. a potential well), the
emission of the proper inerton cloud of one nucleon and the
absorption of the other inerton cloud emitted by another nucleon.
To simplify the investigation, we will consider the
two-dimensional Lagrangian,
\begin{equation}
\label{eq28} {\begin{array}{*{20}c}
 {L = - 2M_{{\kern 1pt}0}
{\kern 1pt} c^{2}\Big\{ {1 - \frac {1}{{2M_{0} {\kern 1pt}
c^{{\kern 1pt}{\kern 1pt} 2}}}} {\kern 1pt} \Big[ {M_{{\kern
1pt}0} \sum\limits_{\iota = 1}^{2} {\left( {\dot {\rho} _{\iota}
^{2} + \rho _{\iota} ^{{\kern 1pt}2} \dot {\Phi} _{\iota} ^{{\kern
1pt} 2}} \right)}
  + m_{{\kern 1pt}0} \sum\limits_{\iota = 1}^{2}
{\left( {\dot {r}_{\iota} ^{2} + r_{\iota} ^{{\kern 1pt}2} \dot
{\phi} _{\iota} ^{{\kern 1pt}2}} \right)}} }
            \hfill                     \\
{\quad {\kern 2pt}  - \frac{{\pi} }{{T}}{\kern 1pt}\sqrt
{m_{{\kern 1pt} 0} M_{{\kern 1pt}0}} {\kern 2pt}
\sum\limits_{\iota = 1}^{2} {f\left( {\rho _{\iota ,{\kern 1pt}
{\kern 1pt} {\kern 1pt} \iota + 1}} \right)} \times  \left( {\rho
_{\iota} \times \left( {\dot {r}_{\iota}  + \dot {r}_{\iota + 1}}
\right) + \rho _{\iota} \Phi _{\iota } \times \left. {( {r_{\iota}
{\kern 1pt} \dot {\phi} _{\iota}  + r_{\iota + 1} {\kern 1pt} \dot
{\phi} _{\iota + 1}} )} \right)} \right.} \hfill
                \\
{{\kern 5pt}  {{{\kern 1pt} {\kern 1pt} {\kern 1pt} \, - \left[ {1
- U_{ +} \left( {|\rho _{{\kern 1pt} 1} + \rho _{{\kern 1pt} 2} |
- 2\lambda _{{\kern 1pt} {\kern 1pt}\rm Com}} \right)} \right]
\times W\left( {|\rho _{{\kern 1pt} 1} - \rho _{{\kern 1pt} 2} |}
\right){\kern 1pt} {\kern 1pt} {\kern 1pt}} \Big]}
\Big\}^{1/2}.\quad\quad \;\;\;\;\;\,\,\quad \,\,} \hfill
\\
\end{array}} \,\,
\end{equation}
Here $M_{0} $ is the mass of a nucleon and $m_{{\kern 0.5pt} 0} =
M_{{\kern 0.5pt} 0} /\sqrt {1 - v ^{2}/c^{2}} $ is the mass of the
nucleon's inerton cloud. In the polar frame of reference $\rho
_{\iota}  $ is the distance to the $\iota $th nucleon reckoned
from the center of inertia of the deuteron that is considered in
the flat model, $\Phi _{\iota}$ is the azimuth of the $\iota $th
nucleon; $\dot {\rho} _{\iota}$ and $\rho _{\iota}  \dot {\Phi
}_{\iota}$ are components of appropriate nucleon velocities.
Similarly, $r_{\iota}  ,\;{\kern 1pt} {\kern 1pt} {\kern 1pt}
{\kern 1pt} r_{\iota} {\kern 1pt} {\kern 1pt} \phi _{\iota}
,\,\,\,\dot {r}_{\iota}$ and $r_{\iota} {\kern 1pt} \dot {\phi}
_{\iota}  $ are corresponding variables of the inerton cloud
accompanying the $\iota $th nucleon. $\pi /T$ is the cyclic
frequency of collisions of the nucleon with its inerton cloud or
with the partner's inerton cloud. The function $f(\rho_\iota) $  correlates the behavior of
the $\iota ${\kern 1pt}th nucleon due to the intervention of the
inerton cloud of the other nucleon. Terms like this ones, $\rho
_{\iota} \left( {\dot {r}_{\iota}  + \dot {r}_{\iota + 1}}
\right)$, mean that the $\iota ${\kern 1pt}th nucleon interacts
simultaneously with its own inerton cloud (emits the $\iota $th
cloud) and with the inerton cloud of the other nucleon (absorbs
the $\left( {\iota + 1} \right)$th cloud). The last term under the
radical sign in the Lagrangian (\ref{eq28}) represents the
interaction induced between two nucleons where their deformation
coats overlap; $W\left( {|\rho _{{\kern 1pt} \iota}  - \rho
_{{\kern 1pt} \iota + 1} |} \right)$ is the potential well that is
formed at the overlapping, which is put constant below, and $U_{
+}  \left( {x} \right)$ is the step function: $U_{ +} \left( {x}
\right) = 0$ if $x \leqslant 0$ and $U_{ +}  \left( {x} \right) =
1$ if $x > 0.$ Note that the expression under the radical sign in
the Lagrangian (\ref{eq28}) is equal to $\sqrt {1 - v ^{2}/c^{2}}$
[13].

The equations of motion obtained from Euler-Lagrange equations
based on the Lagrangian (\ref{eq28}) are too difficult. That is
why let us introduce the following canonical variables, which can
facilitate our consideration in many aspects:
\begin{equation}
\label{eq29} {\begin{array}{*{20}c}
 {Q_{{\kern 1pt} 1{\kern 1pt} \iota}  = r_{\iota}  {\kern 1pt} \dot {\phi
}_{{\kern 1pt} \iota}  \quad  - {\kern 2pt}{\kern 2pt}{\kern 1pt}
\frac{{\pi} }{{2T}}\sqrt {\tfrac{{M_{{\kern 1pt}0} }}{{m_{{\kern
1pt} 0}} }} {\kern 1pt} {\kern 1pt} \rho _{\iota} \Phi
_{\iota},{\kern 1pt} {\kern 1pt} {\kern 1pt} {\kern 1pt} {\kern
1pt} {\kern 1pt}} \hfill
\\
 {Q_{2{\kern 1pt} \iota}  = r_{\iota + 1} {\kern 1pt} \dot {\phi} _{{\kern
1pt} \iota} {\kern 1pt}  - {\kern 5pt} \frac{{\pi} }{{2T}}\sqrt
{\tfrac{{M_{{\kern 1pt}0}} }{{m_{{\kern 1pt} 0}} }} {\kern 1pt}
{\kern 1pt} \rho _{\iota}
\Phi _{\iota},}                  \hfill                     \\
 {q_{{\kern 1pt} 1{\kern 1pt} {\kern 1pt} \iota}  \,\, = \dot {r}_{\iota}
 \quad {\kern 9pt}   - {\kern 6pt} \frac{{\pi} }{{2T}}\sqrt
{\tfrac{{M_{{\kern 1pt}0}} }{{m_{{\kern 1pt} 0}} }} {\kern 1pt}
{\kern 1pt} \rho _{\iota} {\kern 1pt} ,\;\quad \,\,\,} \hfill
                          \\
 {q_{{\kern 1pt} 2{\kern 1pt} {\kern 1pt} \iota}  \, = \dot {r}_{\iota + 1}
\quad  - {\kern 7pt}  \frac{{\pi} }{{2T}}\sqrt {\tfrac{{M_{{\kern
1pt}0}} }{{m_{{\kern 1pt} 0}} }} {\kern 1pt} {\kern 1pt} \rho
_{\iota}
.{\kern 1pt} \,\,\quad \;\,}       \hfill      \\
\end{array}}
\end{equation}

The transition to new variables (\ref{eq29}) changes the
Lagrangian (\ref{eq28}) to the canonical form

\begin{equation}
\label{eq30} {\begin{array}{*{20}c}
 L = - 2M_{{\kern 1pt}0}
 {\kern 1pt} c^{2}\Big\{ 1 - \frac{1}{{2M_{{\kern 1pt}0} {\kern
1pt} c^{2}}} \Big[ M_{{\kern 1pt}0} \sum\limits_{\iota = 1}^{2}
\big( {\dot {\rho} _{\iota} ^{{\kern 1pt} 2} - \omega ^{2} ( {\rho
_{{\kern 1pt} \iota} } ){\kern 1pt} {\kern 1pt} \rho _{\iota}
^{{\kern 1pt} 2}}  \big)
              \hfill   \\
 {\;\quad \quad \quad \quad {\kern 1pt} {\kern 1pt} \,\, + M_{{\kern 1pt}0}
\sum\limits_{\iota = 1}^{2} \big( {\rho _{\iota} ^{{\kern 1pt}
2}{\kern 1pt} \dot {\Phi} _{\iota }^{{\kern 1pt} 2} - \omega ^{2}
( {\rho _{{\kern 1pt} \iota} }) {\kern 1pt} {\kern 1pt} \rho
_{\iota} ^{{\kern 1pt} 2} {\kern 1pt} {\kern 1pt} \Phi _{\iota}
^{{\kern 1pt} 2}} \big)} \hfill
\\
 \quad \quad \quad \quad \quad \, + {\kern 1pt} m_{{\kern 0.5pt}0}
\sum\limits_{\iota = 1}^{2} {\big( {\dot {Q}_{1{\kern 1pt} {\kern
1pt} \iota} ^{{\kern 1pt} 2} + \dot {Q}_{{\kern 1pt}2{\kern 1pt}
{\kern 1pt} \iota} ^{{\kern 1pt}  2} + \dot {q}_{1{\kern 1pt}
{\kern 1pt} \iota }^{\kern 1pt2} {\kern 1pt}  + \dot {q}_{2{\kern
1pt} {\kern 1pt} \iota} ^{{\kern 1pt} 2} {{\kern 1pt}} } \big) }
\hfill
\\
\quad\quad \quad\quad\quad - W \cdot \big[ 1 - U_{ +}  ( {|\rho
_{{\kern 1pt} 1} + \rho _{{\kern 1pt} 2} | - 2\lambda _{{\kern
1pt} {\kern 1pt} \rm Com}}  ) \big] \Big] \Big\}^{1/2}     \hfill
\\
\end{array}}
\end{equation}
where
\begin{equation}
\label{eq31} \omega {\kern 1pt}  \left( {\rho _{{\kern 1pt} \iota}
}  \right) = \frac{{\pi} }{{2T}}{\kern 2pt} f\left( {\rho _{\iota}
}  \right)
\end{equation}
is the effective frequency. Euler-Lagrange equations got from the
Lagrangian (\ref{eq30}) for variables $\rho _{{\kern 1pt} \iota}$
and  $\Phi _{\iota}$  are
\begin{equation}
\label{eq32} \ddot {\rho} _{{\kern 1pt} \iota}  + \omega _{{\kern
1pt} \iota} ^{2} \rho _{{\kern 1pt} \iota}  + \omega _{{\kern 1pt}
\iota}  \frac{{\partial {\kern 1pt} \omega _{\iota} } }{{\partial
\rho _{\iota} } }{\kern 1pt} \rho _{{\kern 1pt} \iota} ^{{\kern
0.5pt} 2} = 0,
\end{equation}
\begin{equation}
\label{eq33} {\begin{array}{*{20}c}
 {\ddot {\Phi} _{\iota}  + \omega _{\iota} ^{2} \Phi _{\iota}  = 0.\quad
\quad}  \hfill \\
 {\quad \quad \,\,\,} \hfill \\
\end{array}}
\end{equation}
If we choose the interaction parameter $f\left( {\rho _{{\kern
1pt} \iota}}\right)$ in the form
\begin{equation}
\label{eq34} f( {\rho _{{\kern 0.3pt}\iota} } ) \equiv f( \rho
_{{\kern 0.3pt}\iota, {\kern 2pt} \iota +1} ) = 1
\end{equation}
we obtain instead of Eq. (\ref{eq32}) the equation
\begin{equation}
\label{eq35} \ddot {\rho} _{{\kern 1pt} \iota}  + \omega _{{\kern
1pt} 0}^{2} \,\rho _{{\kern 1pt} \iota}  = 0
\end{equation}
where $\omega _{0} = 2\pi /T.$

The choice of the parameter $f\left({\rho _{{\kern 1pt} \iota} }
\right)$ in form (\ref{eq34}) allows the harmonic solution to
equations (\ref{eq35}) and (\ref{eq33}),
\begin{equation}
\label{eq36} \rho _{{\kern 1pt} 1} = (-1)^{[t/T]}{\kern 1pt}  \rho _{{\kern 1pt} 0}
\sin(\omega _{\kern 0.6pt 0} {\kern 1pt} t),\quad \quad
\rho _{{\kern 1pt} 2} = (-1)^{[t/T]}{\kern 1pt} \rho _{{\kern 1pt} 0}
\cos\left( {\omega _{\kern 0.6pt 0} {\kern 1pt} t + \varphi}
\right);
\end{equation}
\begin{equation}
\label{eq37} \Phi _{1} = (-1)^{[t/T]}{\kern 1pt} \Phi _{0}  \sin(\omega _{0}
{\kern 1pt} t),\quad \quad \Phi _{2} = (-1)^{[t/T]}{\kern 1pt} \Phi _{0}
\cos\left( {\omega _{0} {\kern 1pt} t + \varphi}
\right).
\end{equation}
The notation $[t/T]$ in expressions (36) and (37)
means an integral part of the integer $t/T$.

Solutions (\ref{eq36}) and (\ref{eq37}) show that nucleons in the
deuteron oscillate along the polar axis and also undergo
rotational oscillations. In other words, the nucleons execute
radial and rotationally oscillatory motions.

In expression (\ref{eq36}) $\rho _{{\kern 1pt} 0} $ means the
maximum removal of the nucleon from the center of inertia of the
deuteron. In other words, the amplitude $\rho _{{\kern 1pt} 0} $
can be associated with the actual radius of the deuteron. If we
suppose that nucleons are quasi-free at the distance of $\rho
_{{\kern 1pt} 0}$, in the first approximation we can prescribe the
frequency $1/T$ to the frequency of collisions of a free nucleon
with its inerton cloud. Note that in this event the de Broglie
wavelength of the nucleon, the said frequency and the nucleon
velocity are linked by the relation $\lambda = v {\kern 1pt}
{\kern 1pt}T$. The value of radius $\rho _{{\kern 1pt} 0}$ can be
deduced by recognizing that $2{\kern 1pt} \pi \rho _{{\kern 1pt}
0} = \lambda$. Since the de Broglie wavelength calculated above
comes out to $\lambda \simeq 1.79 \times 10^{ - 14}$ m, we gain
$\rho _{{\kern 1pt} 0} = \lambda /2\pi \simeq 2.85 \times 10^{ -
15}$ m that is in excess of the radius of the deformation coat
$R_{{\kern 1pt} 0} = 1.32 \times 10^{ - 15}$ m as it must be for
the deuteron.

The calculated value for the radius of deuteron
\begin{equation}
  \label{eq38}
R_{\rm d} = \rho _{{\kern 1pt} 0} = 2.85 \times 10^{ - 15}   \quad
{ \rm m}
\end{equation}
is very close to the estimated effective radius of a coupled
neutron-proton system in the singlet state, i.e. the deuteron
singlet state: $R_{{\kern 0.5pt}\rm d{\kern 1pt} s} = \left( {2.77
\pm 0.05} \right) \times 10^{ - 15}$ m [31]. In the framework of
hadronic mechanics Santilli [32] estimates the value of the
deuteron radius as $R_{\rm d} = 1.41 \times 10^{-15}$ m.

The complete stability of the deuteron depends on many consistent
factors; in particular, the spin factor should also be taken into
account. As has recently been shown [14,15], the particle's spin
introduces an additional kinetic energy in the quantum system
studied. The notion of spin determined in the phase space is
compatible with the proper asymmetric pulsation of the particle in
the real space, which happens forward or backwards (the respective
spin projections $+ 1/2$ or $- 1/2$) relative to the particle
vector velocity [14] (see also Ref. [7]). Therefore the stable
triplet state that is realized in the deuteron should imply that
the correction to the kinetic energy of the nucleons on the side
of spin increases their total energy during collisions inside the
deuteron (hence the Lagrangians (\ref{eq28}) and (\ref{eq30})
should be complemented by additional terms). The increase in
energy helps them move closer to one another, which in turn should
strengthen the overlapping of the nucleons' deformation coats,
i.e. this will enhance the attraction of the nucleons.

Since we touch upon the problem of the spin, the following points
need to be made. It is generally recognized that nuclear forces
depend on spin. In the case of the nucleon, spin-1/2 should
correspond to the nucleon's proper pulsations, as in the case with
a canonical particle [14]. Namely, anisotropic pulsations of the
nuclear core (with radius $R_{{\kern 1pt}\rm  c}$), which
represents a hard ball holding quarks, should be associated with
spin-1/2 of the nucleon. The nucleon's cloud of inertons
transports all the properties of the core. Therefore, inertons,
oscillating around the moving nucleons, should carry fragments of
the core pulsations, i.e. the nucleon core emitting inertons along
a section $\lambda /2$ passes fragments of its anisotropy to
inertons and gradually loses the anisotropy deformation (spin).
During the next section $\lambda /2$ the nucleon core absorbs
inertons, which gives back the anisotropy pulsation to the core,
and so on. So, the inerton cloud also transports, among other
properties of the nucleon core, its spin polarization. Owing to
the nucleon-nucleon interaction, which occurs by means of the
nucleons' inertons, the spin component should also be imposed upon
nuclear forces (in particular, at scattering of neutrons/protons
by molecular compounds, especially those, which include molecular
or atomic hydrogen).

In the limiting case of zero energy of the relative motion of
neutron and proton the cross-section for their scattering $\sigma$
is defined by the value of the binding energy of deuteron $\Delta
E = 2.22$ MeV. Since the actual size of a moving nucleon lies in
in the range of the deformation coat $R_{{\kern 1pt} 0}$ and the
radius of the inerton cloud $R = \tfrac{{1}}{{2}}\sqrt {\Lambda
^{2} - \lambda ^{2}},$ the value of $\sigma$ should satisfy the
inequality $\pi R_{\kern 1pt 0}^{\kern 1pt 2} < \sigma < \tfrac
{\pi}{2} \left( {\Lambda ^{2} - \lambda ^{2}} \right)$. The
effective value of $\sigma $ can be deduced from the scheme shown
in Figure 1. Indeed, the cloud of inertons spreads on $\lambda$
along the nucleon path and on the distance of $\tfrac 12 \sqrt
{\Lambda ^{2} - \lambda ^{2}}$ around the nucleon in the
transverse directions. The cross-section of such a spindle-shaped
body is
$$
\sigma = \tfrac 12 {\kern 1pt}\lambda {\kern 1pt}\sqrt
{\Lambda ^{2} - \lambda ^{2}} = 2.33 \times 10^{ - 27} \ {\rm m}^2
$$
(we use here numerical values of the corresponding parameters
calculated above: $\lambda \simeq 1.79 \times 10^{ - 14}$ m and
$\Lambda \simeq 2.6 \times 10^{ - 13}$ m). This estimate has been
obtained under the assumption that nucleons are scattered across
the space. The estimate correlates well with the experimental
value of the cross-section $\sigma _{\rm exper} = 2.05 \times 10^{
- 27}$ m$^{2}$ [33].

\section*{5. Nucleus as a cluster of nucleons}

\hspace*{\parindent} Since nuclei consist of protons and neutrons,
it should be the reason for such a combination of nucleons. Let us
consider the nucleus stability reasoning from the statistical
description of the system of a great quantity of interacting
protons and neutrons.

An interesting approach to the statistical description of the
system of interacting particles, which makes allowance for spatial
nonhomogeneous states of particles in the system studied, was first
proposed in Ref. [34]. However, if the inverse operator of the
interaction energy cannot be determined, a different method should
be applied. It will make it possible to take into account a
possible nonhomogeneous particle distribution. In paper [35]
systems of interacting particles were treated from the same
standpoint. Nevertheless, the number of variables describing the
systems in question was reduced because of introducing a new
canonical variable into equations of equilibrium, which
characterized a nascent nonhomogeneous state (i.e. a cluster). So
the nonhomogeneous state automatically arose as the logical
consequence of the behavior of particles. It should be noted that
the major peculiarity of the said approach is that the pair
potential has been broken into two components, namely, the
approach allows the isolation of attractive and repulsive
components with the further study of a possible equilibrium
nonhomogeneous state of particles.

Before applying the aforementioned statistical approach, we shall
first clarify the structure (or nature) of the pair potential,
which acts between nucleons, and then subdivide the potential into
attractive and repulsive components.

\subsection*{\textbf{\textit{5.1. Hamiltonian of two kinds of
interacting particles} }}

\hspace*{\parindent} We shall start from the construction of the
Hamiltonian for a system of two kinds of interacting particles,
namely, neutrons and protons. Let particles, i.e. nucleons, form a
3D lattice and let $n = \{ 0,\,1\} $ be the filling number of the
\textit{s}th lattice knot. The energy for such a system can be
written in the form [36]
\begin{equation}
\label{eq39} {\begin{array}{*{20}c} {H = \sum\limits_{\bf r,{\kern
1pt} {\kern 1pt} {\bf r}^{\kern 1pt \prime}} {\upsilon _{\rm
{\kern 1pt} n{\kern 1pt} n} \left( {\bf r,{\kern 1pt} {\kern 1pt}
{\bf r}^{\kern 2pt \prime}} \right){\kern 1pt} {\kern 1pt}} c_{\rm
n} \left( {\bf r} \right){\kern 1pt} {\kern 1pt} c_{\rm n} \left(
{{\bf r}^{\kern 2pt \prime}} \right) + 2\sum\limits_{\bf r,{\kern
1pt} {\kern 1pt} {\bf r}^{\kern 2pt \prime}} {\upsilon _{\rm
{\kern 1pt} n{\kern 1pt} p} \left( {{\bf r},{\kern 1pt} {\kern
1pt} {\bf r}^{\kern 1pt \prime}} \right){\kern 1pt} {\kern 1pt}}
c_{\rm n} \left( {\bf r} \right){\kern 1pt} {\kern 1pt} c_{\rm p}
\left( {{\bf r}^{\kern 2pt \prime}} \right)} \hfill
\\
 {\quad \quad + \sum\limits_{\bf r, {\kern 1pt} {\bf r}^{\kern 2pt
\prime}} {\upsilon _{\rm {\kern 1pt} p{\kern 1pt} p} \left( {\bf
r,{\kern 1pt} {\kern 1pt} {\bf r}^{\kern 1pt \prime}}
\right){\kern 1pt} {\kern 1pt}} c_{\rm p} \left( {\bf r}
\right){\kern 1pt} {\kern 1pt} c_{\rm p} \left( {{\bf r}^{\kern
1pt \prime}} \right)\quad \quad \quad \quad \quad \quad \quad
\quad} \hfill \\
\end{array}}
\end{equation}

\noindent where $\upsilon _{{\kern 0.3pt}i j}$ is the interaction
potential of nucleons of two kinds, $i,{\kern 1pt} {\kern 1pt}
{\kern 1pt} j ={\rm  n},\,{\rm p}$ (n stands the neutron and p
stands for the proton). They occupy knots in Ising's lattice
described by the radius vectors $\bf r$ and ${\bf r}^{{\kern 1pt}
\prime}$ and $c_{{\kern 1pt} i}({\bf r})=\{0,{\kern 2pt}1\}$
are the random functions, which satisfy condition
\begin{equation}
\label{eq40} c_{\rm n} \left( {\bf r} \right) + c_{\rm p} \left(
{\bf r} \right) = 1.
\end{equation}

The Hamiltonian (\ref{eq39}) can be written as follows
\begin{equation}
\label{eq41} H = H_{0} + \tfrac{1}{2} \sum\limits_{\bf r,\, {{\bf
r}^{\kern 1pt \prime}}} {\upsilon \left( {{\bf r},\,{\kern 1pt}
{{\bf r}}^{\kern 1pt \prime}} \right)} {\kern 1pt} {\kern 1pt}
{\kern 1pt} c_{\rm n} \left( {{\bf r}} \right){\kern 1pt} {\kern
1pt} {\kern 1pt} c_{\rm n} \left( {{{\bf r}}^{\kern 1pt \prime}}
\right)
\end{equation}
where
\begin{equation}
\label{eq42} H_{0} = \tfrac{{1}}{{2}}\sum\limits_{{\bf r}} {\left[
{ 1 - 2{\kern 1pt}c_{\rm n} \left( {{\bf r}} \right)} \right]}
\sum\limits_{{{\bf r}}^{\kern 1pt \prime}} {\upsilon _{\rm p{\kern
1pt} p} \left( {{\bf r},\,{{\bf r}}^{\kern 2pt \prime}} \right)} +
\sum\limits_{{\bf r}} {c_{\rm n} \left( {\bf r} \right){\kern 1pt} }
\sum\limits_{{{\bf r}}^{\kern 1pt \prime}} {\upsilon _{\rm p{\kern
1pt} n} \left( {{\bf r},\,{{\bf r}}^{\kern 1pt \prime}} \right)} ,
\end{equation}
\begin{equation}
\label{eq43} \upsilon \left( {{\bf r}, \,{{\bf r}}^{\kern 1pt
\prime}} \right) = \upsilon _{\rm n{\kern 1pt} n} \left( {{\bf r},
\, {{\bf r}}^{\kern 1pt \prime}} \right) + \upsilon _{\rm p{\kern
1pt} p} \left( {{\bf r}, \, {{\bf r}}^{\kern 1pt \prime}} \right)
- 2{\kern 1pt}\upsilon _{\rm p{\kern 1pt} n} \left( {{\bf r},
\,{{\bf r}}^{\kern 1pt \prime}} \right).
\end{equation}

Let us rewrite the Hamiltonian (\ref{eq41}) in the form
\begin{equation}
\label{eq44} H = H_{0} - \tfrac{{1}}{{2}}\sum\limits_{{\bf
r},\,{\kern 1pt} {{\bf r}}^{\kern 1pt \prime}} {V_{{\bf r}{\kern
1pt} {{\bf r}}^{\kern 1pt \prime}}}  {\kern 1pt} {\kern 1pt}
c{\kern 1pt} {\kern 1pt} \left( {{\bf r}} \right){\kern 1pt} c
\left( {{{\bf r}}^{\kern 1pt \prime}} \right) +
\tfrac{{1}}{{2}}\sum\limits_{{\bf r},\,{\kern 1pt} {{\bf
r}}^{\kern 1pt \prime}} {U_{{\bf r}{\kern 1pt} {{\bf r}}^{\kern
1pt \prime}}} {\kern 1pt} {\kern 1pt} c {\kern 1pt} \left( {{\bf
r}} \right){\kern 1pt} c  \left( {{{\bf r}}^{\kern 1pt \prime}}
\right)
\end{equation}
where the index p is omitted at the function $c{\kern 1pt} \left(
{r} \right)$ and the following designations are introduced:
\begin{equation}
\label{eq45} V_{{\bf r}{\kern 1pt} {{\bf r}}^{\kern 1pt \prime}} =
\upsilon _{\rm p{\kern 1pt} n} \left( {{\bf r}, {\kern 2pt} {{\bf
r}}^{\kern 2 pt \prime}} \right),
\end{equation}
\begin{equation}
\label{eq46} U_{{\bf r}{\kern 1pt} {{\bf r}}^{\kern 1 pt \prime}}
= \tfrac{{1}}{{2}}{\kern 1pt}
 \left[ {\upsilon _{\rm n{\kern 1pt} n} \left( {{\bf r},\,
{{\bf r}}^{\kern 1pt \prime}} \right){\kern 1pt} \,{\kern 1pt} +
\,\,\upsilon _{\rm p{\kern 1pt} p} \left( {{\bf r},{\kern 2pt}
{{\bf r}}^{\kern 1pt \prime}} \right)} \right].
\end{equation}

If the potentials $V_{{\bf r}{\kern 1pt} {{\bf r}}^{\kern 2 pt
\prime}} $, $U_{{\bf r}{\kern 1pt} {{\bf r}}^{\kern 2 pt \prime}}
> 0$, the second term in the right-hand side of the
Hamiltonian (\ref{eq42}) corresponds to the effective attraction
(45) and the third term conforms to the effective repulsion (46).
This allows one to represent the Hamiltonian (\ref{eq42}) in the
form that is typical for the model of ordered particles, which is
characterized by a certain nonzero order parameter,
\begin{equation}
\label{eq47} H\left( n \right) = \sum\limits_s E_s {\kern 1pt} n_s
- \tfrac 12 \sum\limits_{s,{\kern 1pt} s^{{\kern 1pt}\prime}}
{\kern 1pt} V_{s{\kern 1pt} s^{{\kern 1pt}\prime} }  {\kern 1pt}
n_s {\kern 1pt} n_{s^{{\kern 1pt}\prime} }
 + \tfrac 12 \sum\limits_{s,{\kern 1pt}
s^{{\kern 1pt}\prime} } {\kern 1pt} U_{s{\kern 1pt} s^{{\kern
1pt}\prime}} {\kern 1pt} n_{s} {\kern 1pt} n_ {s^{{\kern
1pt}\prime}}.
\end{equation}
Here $E_{s} $ is the additive part of the particle energy (the
kinetic energy) in the \textit{s}th state. The main point of our
approach is the initial separation of the total nucleon potential
into two terms: the repulsion and attraction components. So, in
the Hamiltonian (\ref{eq47}) the potential $V_{s{\kern 1pt}
s^{\kern 0.5pt \prime }}$ represents the paired energy of
attraction and the potential $U_{s{\kern 1pt} s^{\kern 0.5pt
\prime }}$ is the paired energy of repulsion. The potentials take
into account the effective paired interaction between nucleons
located in states $s$ and $s^{\kern 1.5pt \prime}$. The filling
numbers $n_s$ can run only two meanings: 1 (the $s$th knot is
occupied in the model lattice studied) or 0 (the $s$th knot is not
occupied in the model lattice studied). The signs before positive
functions $V_{s{\kern 1pt} {s}^{\kern 0.5pt \prime }}$ and
$U_{s{\kern 1pt} s^{\kern 0.5pt \prime }}$ in the Hamiltonian
(\ref{eq47}) directly specify proper signs of attraction (minus)
and repulsion (plus).

\subsection*{\textbf{\textit{5.2. Statistical mechanical approach}}}

\hspace*{\parindent} The statistical sum of the system under
consideration
\begin{equation}
\label{eq48} Z = \sum\limits_{\{ n\}}  {\exp {\kern 1pt} \left( {
- H\left( {n} \right)/k_{\rm B} \Theta}  \right)}
\end{equation}

\noindent
can be presented in the field form
\begin{eqnarray}
Z &=& \int\limits_{ - \infty}
^{\infty}  D{\kern 1pt} {\kern 1pt} \phi \int\limits_{ -
\infty} ^{\infty}  D {\kern 1pt} {\kern 1pt} \psi
\sum\limits_{ \{ n \} }   \exp \Big[  - \sum\limits_s   E_s {\kern
1pt} n_s + \sum\limits_s \big( \psi_s + i{\kern 1pt}
\phi_s \big) {\kern 1pt} {\kern 1pt} n_s
\nonumber        \\
&& \qquad\qquad\qquad\qquad\qquad\quad
- \frac{1}{2} \sum\limits_{s,{\kern 1pt} s^{{\kern 0.5 pt} \prime}}
\big( {\kern 1pt} {\tilde V}_{s{\kern 1pt} s^{\kern 1pt
\prime}}^{ - 1} {\kern 1pt} \psi_s {\kern 1pt} \psi_{s^{{\kern
0.5pt} \prime}}  + {\tilde U}_{s{\kern 1pt} s^{\kern0.5pt \prime}}^{ - 1}
{\kern 1pt} {\kern 1pt} \phi _s {\kern 1pt} {\kern 1pt}
\phi_{s^{\kern 0.5pt \prime}} {\kern 1pt} \big)
\Big]
\label{eq49}
\end{eqnarray}

\noindent
due to the following representation known from the theory of Gauss integrals
\begin{equation}
\label{eq50}
\begin{array}{l}
 \exp\left( {\tfrac{{\beta} }{{2}}\sum\limits_{s, {\kern 0.6pt}s^{\kern 0.5pt \prime}}
 {W_{ss^{\kern 0.5pt \prime}} {\kern 1pt}
n_{s} {\kern 1pt} n_{s^{\kern 0.5pt \prime}}} }  \right)  = {\rm
Re}\int\limits_{ - \infty} ^{\infty} {{ D}{\kern 1pt} {\kern 1pt}
\chi}  \exp {\kern 1pt} \left( {\beta \sum\limits_{s} {n_{s} \chi
_{s}}  - \tfrac{{1}}{{2}}\sum\limits_{s,{\kern 1.5 pt}s^{\kern 0.5pt \prime}}
{W_{ss^{\kern 0.5pt \prime}}^{ -
1}}  {\kern 1pt} \chi _{s} {\kern 1pt}
\chi _{s^{\kern 1.5 pt \prime}}}  \right), \\
 \qquad \quad \quad \ \ \
 W_{ss^{\kern 0.5pt \prime \prime}}^{ - 1} {\kern 1pt}
 W_{s^{\kern 0.5pt \prime \prime}
 {s^{\kern 0.5pt \prime}}} = \delta _{ss^{\kern 0.5pt \prime}} \\
 \end{array}
\end{equation}
where $D\chi \equiv \prod\nolimits_{s} {\sqrt {\det ||W_{s{\kern
1pt} {s}'} ||}}  {\kern 1pt} {\kern 1pt} {\kern 1pt} 2\pi {\kern
1pt} {d}\chi _{s} $ implies the functional integration with
respect to the field $\chi $, $\beta ^{{\kern 1pt} 2} = \pm 1$ in
relation to the sign of interaction (+1 for attraction and $ - 1$
for repulsion). The dimensionless energy parameters $\tilde
{V}_{s{s}'} = V_{s s'}/k_{\rm B} \Theta $, $\tilde {U}_{s s'} = U_{s s'}/k_{\rm
B} \Theta $ and $\tilde {E}_{s} = E_s /k_{\rm B} \Theta $, where
$\Theta $ is the temperature, are introduced into expression
(\ref{eq48}). Passing to the canonical ensemble of \textit{N}
particles and summing over $n_{s} $ (note $\sum\nolimits_{s}
{n_{\kern 0.3pt s} = N} $), we will get instead of (\ref{eq49})
for the case of the Fermi statistics [35]
\begin{equation}
\label{eq51} Z = {\rm Re}\frac {1}{2\pi i} \int  D{\kern 1pt}
{\kern 1pt} \phi \int D{\kern 1pt} {\kern 1pt} \psi {\kern 1pt}
{\kern 1pt} \oint d{\kern 1pt} z {\kern 2pt}  \exp \left[ S \left(
{\phi ,\,\psi ,\,z}\right) \right]
\end{equation}

\noindent where
\begin{equation}
\label{eq52} {\begin{array}{*{20}c}
 {S = \sum\limits_{s} {\left\{ { - \tfrac{{1}}{{2}}\sum\limits_{{s}'}
{\left( {\,\tilde {U}_{s{s}'}^{ - 1} {\kern 1pt} \phi _{s} {\kern
1pt} \phi _{{s}'} + \tilde {V}_{s{s}'}^{ - 1} {\kern 1pt} \psi
_{s} {\kern 1pt} \psi _{{s}'}}  \right)} \;\;\;\;\;\;\quad}
\right.}} \hfill \\
 {\quad \quad \quad \ \ \ \  \left. { + \ln\left| {{\kern 1pt} 1 + \frac{{1}}{{z}}
{\kern 1pt}\exp \left( - {\tilde E}_s  + \psi _s \right) \cos\phi
_{s}} \right|} \right\} + \left( {N - 1} \right)\ln z.}
\hfill \\
\end{array}}
\end{equation}

Let us set $z = \xi + i{\kern 1pt} \zeta$ and consider the action
$S$ on the transit path passing through the saddle-point with a
fixed imaginable variable ${\rm Im}{\kern 1pt} z = \zeta _{{\kern
1pt} 0} $. In this case, it stands to reason that the action
\textit{S}, similarly to quantum field theory, is the variational
functional that depends on three variables, namely, fields $\phi
_{{\kern 1pt} s}$ and $\psi _{{\kern 1pt} s}$, and the fugacity
$\xi = \exp( - \mu /k_{\rm B} \Theta)$ where $\mu$ is the chemical
potential. The extremum of the functional must be realized at
solutions of the equations $\delta {\kern 1pt} S/\delta {\kern
1pt} \phi _{{\kern 1pt} s} = 0, \  \delta {\kern 1pt} S/\delta
{\kern 1pt} \psi _{{\kern 1pt} s} = 0\quad$ and $\delta {\kern 1pt}
S/\delta {\kern 1pt} \xi = 0$. These equations appear as follows:
\begin{equation}
\label{eq53} \sum\limits_{s'} {\tilde {U}{\kern 1pt} _{s{\kern 1pt}
s^{\kern 0.5pt \prime}}^{ - 1}} {\kern 1pt} {\kern 1pt} \phi
_{s^{\kern 0.5pt \prime}} {\kern 1pt} = - \frac{{2{\kern 1pt}
{\kern 1pt} \exp( - {\tilde E}_s + \psi _s) {\kern 1pt}\sin\phi
_{s}} }{{\xi - \exp( - {\tilde E}_s + \psi _s) {\kern 1pt} \cos\phi
_{s}}},
\end{equation}
\begin{equation}
\label{eq54}
 \sum\limits_{s'} {\tilde {V}{\kern 1pt} _{s{\kern 1pt}
s^{\kern 0.5pt \prime}}^{ - 1}} {\kern 1pt} {\kern 1pt} \psi
_{s^{\kern 0.5pt \prime}} {\kern 1pt } = {\kern 1pt }
\frac{{2{\kern 1pt} {\kern 1pt} \exp( - {\tilde E}_s + \psi
_s){\kern 1pt} \cos\phi _{s}} }{{\xi - \exp( - {\tilde E}_s + \psi
_s) {\kern 1pt}\cos\phi _{s}} }{\kern 1pt},
\end{equation}
\begin{equation}
\label{eq55}
 \sum\limits_s
\frac { \exp( - {\tilde E}_s + \psi _s){\kern 1pt} \cos\phi
_s }{\xi - \exp( - {\tilde E}_s + \psi _s)
{\kern 1pt} \cos\phi _s}  = N - 1.
\end{equation}

If we introduce denotation
\begin{equation}
\label{eq56} A_{{\kern 1pt} s} = \frac{ \exp( - {\tilde E}_s + \psi
_s) \cos\phi _{s} }{{\xi - \exp( - {\tilde E}_s + \psi _s) \cos\phi
_{s}} },
\end{equation}

\noindent we will see from Eqs. (53)-(55) that the sum
$\sum\nolimits_{s} {A_{{\kern 1pt} s}}  + 1$ is exactly equal to
the total number of particles $N$ in the system studied, i.e.
\begin{equation}
\label{eq57} \sum\limits_{s} {A_{{\kern 1pt} s}}  = N - 1.
\end{equation}

So it follows from Eq. (53)-(55) that the parameter $A_{{\kern
1pt} s}$, which combines variables $\phi _{s} $, $\psi _{s}$, and
$\xi$, is a typical variable that characterizes the number of
particles contained in the \textit{s}th cluster.

The action (\ref{eq52}) can be rewritten in terms of variables
$A_{{\kern 1pt} s} $ and $\xi $ as follows [35]:
\begin{equation}
\label{eq58} {\begin{array}{*{20}c} {S = - \frac 12
\sum\limits_{s,{\kern 1pt} {s}'} {{\kern 1pt} \tilde {V}_{s{\kern
1pt} {s}'}}  {\kern 1pt} A_{{\kern 1pt} s^{\kern 0.4pt \prime}}
{\kern 1pt} A_{{\kern 1pt} s} + \sum\limits_{s} {\ln|1 + A_{{\kern
1pt} s} |} + \left( {N - 1} \right) \ln \xi - \frac 12
\sum\limits_{s,{\kern 1pt} {s}'} {\tilde {U}_{s{\kern 1pt} s^{\kern
0.4pt \prime}}} } \hfill
\\
 {\quad\quad\quad \times \left\{ {\xi ^{{\kern 1pt}
- 2}{\kern 2pt} \exp      \left(     - 2\tilde {E}_{s^{\kern 0.5pt
\prime}} {\kern 1pt} + 2\sum\nolimits_{s^{{\kern 0.5pt} \prime
\prime}} {\tilde V}_{s^{\kern 0.5pt \prime} s^{\kern
0.5pt \prime \prime}} {\kern 1pt} A_{s^{\prime \prime}} \right)
\left( {1 + A_{{\kern 1pt} s^{\kern 0.5pt \prime}}} \right)^{2} -
A_{{\kern 1pt} {s^{\kern 0.5pt \prime}}}^{2}} \right\}^{1/2}} \hfill \\
{ \quad\quad\quad \times \left\{ {\xi ^{{\kern 1pt} - 2}{\kern
2pt} {\kern 2pt} \exp  \left( - 2{\tilde E}_s {\kern 1pt} + {\kern
1pt} 2 \sum\nolimits_{s^{{\kern 0.5pt} \prime \prime}} {\kern
1pt}{\tilde V}_{s{\kern 1pt} s^{{\kern 0.5pt} \prime
\prime}} {\kern 1pt} A_{s^{\prime \prime}}  \right) \left( 1 +
A_{{\kern 1pt} s} \right)^{2} {\kern 1pt} - {\kern 1pt} A_{{\kern
1pt} s}^{2}} \right\}^{1/2}}. \hfill
\\
\end{array}}
\end{equation}

We can simplify our consideration if we deem that all nucleons are
distributed by clusters and each cluster includes the same number
\textit{A }of nucleons. This changes Eq. (\ref{eq55}) to the
following:
\begin{equation}
\label{eq59} A{\kern 1pt} K = N - 1
\end{equation}
where the combined variable \textit{A} now represents a number of
nucleons in one cluster of the system of \textit{N} nucleons. Now
we can pass to continual variables into the action (\ref{eq58}).
Assuming that the density of particles is different from zero only
in the cluster volume $\mathcal{V} = 4\pi g^{3}A/3$ where $g$ is
the distance between nucleons in a cluster (i.e. the lattice
constant) and therefore $g{\kern 1pt} A^{1/3}$ is the radius of
the cluster, we can strongly facilitate the form of the action
(\ref{eq58}) [35]
\begin{eqnarray}
\label{eq60} S &=& \tfrac 12 {\kern 1pt} K \times \left\{ {\left(
{a - b} \right){\kern 1pt} {\kern 1pt} A^{\kern 1pt 2} - \frac{{1
- \langle n\rangle} }{{\langle n\rangle} }{\kern 1pt} {\kern 1pt}
{\kern 1pt} \exp \left( {\kern 1pt}2{\kern 1pt}b{\kern 1pt} A
\right) {\kern 1pt} \left( {A^{ - 1} + 1} \right)^{2}{\kern
1pt}  - \ln \left( {A + 1} \right)} \right\}  \nonumber    \\
&& + \left( {N - 1} \right) \ln \xi
\end{eqnarray}
where the following functions are introduced:
\begin{equation}
\label{eq61} a = 3 {\kern 1pt} \int_{1}^{A^{1/3}} {\tilde {U}(
{g{\kern 1pt} x}){\kern 1pt} {\kern 1pt} {\kern 1pt} x} {\kern
1pt} ^{2}{\kern 1pt} d{\kern 1pt} x,
\end{equation}
\begin{equation}
\label{eq62} b = 3 {\kern 1pt} \int_{1}^{A^{1/3}} {\tilde {V}(
{g{\kern 1pt}x}) {\kern 1pt} {\kern 1pt} {\kern 1pt} x} {\kern
1pt} ^{2}{\kern 1pt} d{\kern 1pt} x
\end{equation}
(recall that tilde over the symbols means division by the factor
$k_{\rm B} \Theta$). Here, in expression (\ref{eq60}), we have
substituted the function $\xi^{-2} \exp(- 2\tilde {E})$
for the mean number of Fermions $\langle n\rangle$ in
one-particle state, i.e.
  $$
\xi^{-2}{\kern 1pt} \exp( - 2\tilde {E}) \equiv \exp(
- 2\tilde {E} + \tilde {\mu})  = \left( {1 - \langle n\rangle}
\right){\kern 1pt} {\kern 1pt} /{\kern 1pt} {\kern 1pt} \langle
n\rangle ;
  $$
besides, we have introduced the dimensionless variable $x$ under
integrals in expressions (61) and (62).

\subsection*{\textbf{\textit{5.3. Realistic potentials}} }

\hspace*{\parindent} Nucleons fill shells in a nucleus and move in
the field of the same potential. The potential shape is formed in
a range of the space that features an additional deformation of
superparticles, i.e. the attraction, owing to i) the overlapping
of deformation coats of nucleons and ii) the overlapping of
inerton clouds of nucleons, which carry the space deformation,
i.e. mass, as well. It is evident that in the field of the
potential nucleons move along their proper trajectories and each
of the nucleons possesses its own energy, the moment of momentum
and spin. If we project such a dynamic behavior of nucleons on a
model lattice, we arrive at the pattern in which each nucleon
occupies its own knot. Let us leave room for the interaction
between nucleons located in the said knots.

Although the mass distribution in the inerton cloud of a nucleon
is governed by law (\ref{eq23}),  the paired interaction on the
scale $r \ll \Lambda $ can be simulated according to harmonic law
(see also Ref. [21]): Nucleons having the same energy and momentum
shall behavior like elastic balls. In other words, nucleons'
inerton clouds will elastically interact each other in the range
$r < \Lambda $. Therefore, we can reason that in the model lattice
of a system of the huge number $N$ of nucleons the attraction
between nucleons, associated with the space deformation of the
kinds i) and ii) mentioned above, can be presented in the form of
two terms, namely,
\begin{equation}
\label{eq63} V = - {\kern 1pt} {\kern 1pt} \hbar {\kern 1pt}
{\kern 1pt} \omega_1{\kern 2pt} \frac{{\Delta R}}{{R_{{\kern 1pt}
0}} }\,\,\, +  \ \   \tfrac {1}{2} {\kern 1pt}{\kern 1pt}\gamma
{\kern 1pt} {\kern 1pt} r^{{\kern 0.5pt}2}
\end{equation}
where ${\kern 1pt} {\kern 1pt} \hbar {\kern 1pt}{\kern 1pt} \omega
_{ 1} $ is the total energy of  nucleon (17); $r$ is the
difference between positions of a pair of nucleons in the lattice
studied and  $\gamma $ is the elasticity/force constant. Such
harmonic potential is often employed in nuclear physics, though so
far its origin has remained complete unclear. The potential (63)
should be identified with the effective attractive potential
(\ref{eq45}), which enters in the function $b$ (62).

The electromagnetic interaction, i.e. repulsion, which happens
between protons, is realized via protons' deformation coats.
According our results [22,11,8] the electromagnetic polarization,
which is a peculiar surface fractality of cells of the
tessellattice, is imposed upon the mass deformation (i.e. the
volume fractality change of a cell) that places the leading role
in the deformation of the space. Therefore we shall preserve law
(\ref{eq23}) for the description of the charge drop in the
polarized nucleon's coat with distance.  If we denote the
repulsion between protons associated with their electromagnetic
interaction as
\begin{equation}
  \label{64}
U_{\rm el. - magn.} = \frac{{1}}{{4{\kern 1pt} \pi {\kern 1pt}
\varepsilon _{\kern 0.2pt 0}} }{\kern 2pt} \frac {e^{{\kern
1pt}2}}{r},
\end{equation}
we can rewrite the potential of effective repulsive (46) as
follows
\begin{equation}
\label{eq65} U = \tfrac{{1}}{{2}}\left[ {\left( { - {\kern 1pt}
{\kern 1pt} \hbar {\kern 1.5pt} {\kern 1pt} \omega_1 {\kern 1pt}
\frac{{\Delta R}}{{R_{{\kern 1pt} 0}} }\,\,\, +
\,\,\,\tfrac{{1}}{{2}}{\kern 1pt}{\kern 1pt}\gamma {\kern 1pt}
{\kern 1pt} r^{\kern 0.5pt 2}} \right) + \left( { - {\kern 1pt}
{\kern 1pt} \hbar {\kern 1pt} {\kern 1.5pt} \omega_1 {\kern 1pt}
\frac{{\Delta R}}{{R_{{\kern 1pt} 0}} }\,\, + {\kern 1pt} {\kern
1pt} \frac{1}{{4{\kern 1pt} \pi {\kern 1pt} \varepsilon _{{\kern
0.2pt} 0}} }{\kern 2pt} \frac {e^{{\kern 1pt}2}}{r}  } \right)}
\right].
\end{equation}

Calculating functions $a$ (\ref{eq61}) and $b$ (\ref{eq62}) we get
instead of action (60)
\begin{eqnarray}
\label{eq66}
 && S \cong \tfrac{1}{2} {\kern 1pt} K \times \left\{
 \Big( {\frac{{3{\kern 1pt} {\kern 1pt} e^{\kern 0.4pt 2}A^{\kern 1pt
2/3}}}{{16{\kern 1pt} {\kern 1pt} \pi {\kern 1pt} \varepsilon _{0}
{\kern 1pt} g{\kern 2pt} k_{\rm B} \Theta} }{\kern 1pt} {\kern
1pt} {\kern 1pt} {\kern 1pt} - {\kern 1pt} {\kern 1pt} {\kern 1pt}
{\kern 1pt} \frac{{3{\kern 1pt} {\kern 1pt} \gamma {\kern 1pt}
{\kern 1pt} g^{\kern 0.4pt 2}A^{{\kern 1pt} 5/3}}}{{20{\kern 1pt}
{\kern 1pt} {\kern 1pt} k_{\rm B} \Theta} } \Big) A^{\kern 1pt 2}-
\ln {\kern 1pt} {A}} \right\} \nonumber
\\
 && \quad\quad  +  \ O\left( {1 - \langle n\rangle}
\right) + \left( {N - 1} \right) \ln \xi
\end{eqnarray}
where the function $O\left( {1 - \langle n\rangle} \right)$
denotes small terms, which can be neglected owing to the strong
degeneration of nucleons, i.e. due to the inequality $1 - \langle
n\rangle \ll 1$.

The minimum of action (66) is reached at the solution of the
equation $\partial S/\partial A = 0$ (if the inequality $\partial
^{{\kern 1pt} 2}S/\partial A^{2} > 0$ holds). With the
approximation $A \gg 1$ the corresponding solution is
\begin{eqnarray}
\label{eq67} A = \frac{5{\kern 2pt}e^{{\kern 0.4pt}2}}{ 4{\kern
1pt}\pi \varepsilon _{0} {\kern 1pt} \gamma {\kern 1pt} {\kern
1pt} g^{{\kern 0.5pt} 3}} \equiv \frac{5}{3}{\kern 1pt} {\kern
1pt} {\kern 1pt} {\kern 1pt} \frac{{e^{2}\rho _{{\kern 0.4pt}\rm
nucl}} }{{\varepsilon _{0} {\kern 1pt} \gamma} }
\end{eqnarray}
where $\rho _{\rm nucl} $ is the density of the nuclear matter;
below we set $\rho _{\rm nucl} = 1.68\times 10^{44}$ m$^{-3}$.
This solution shows that the number of nucleons $A$ in a nucleus
is set for a given number of charged protons whose repulsion is
balanced out by the mean inerton field of all the $A$ nucleons.

A solution similar to the result (67) has been obtained in our
paper [35] with the use of the Yukawa potential. However, the
result obtained is rather formal, though it has taken into account
the deviation of the potential from the pure Coulomb law, as
expression (23) prescribes.

\subsection*{\textbf{\textit{5.4. Nuclear instability}}}

\hspace*{\parindent} In Table 1 based on result (\ref{eq67}) some
major parameters of a nucleus with $A$ nucleons are presented. The
value $\hbar {\kern 1pt}{\kern 1pt} \omega $ represents the
difference between equidistant levels of energy in the solution of
the corresponding oscillatory problem. We see from Table 1 that,
with \textit{A} increasing it is the elasticity constant $\gamma $
of the inerton field that mostly changes, which in turn leads to a
dramatic decrease in $\hbar {\kern 1pt} {\kern 1pt}\omega $. In
fact, when $\hbar {\kern 1pt} {\kern 1pt}\omega $ falls to the
critical value 0.75 MeV, which corresponds to the Coulomb
repulsion between protons, the cluster (i.e. the nucleus) begins
to decompose.

\bigskip

\textbf{Table 1.} Estimation of some parameters

\newcommand{\PreserveBackslash}[1]{\let\temp=\\#1\let\\=\temp}
\let\PBS=\PreserveBackslash

\begin{longtable}
{|p{22pt}|p{60pt}|p{60pt}|p{28pt}|p{64pt}|p{64pt}|p{26pt}|} a & a
& a & a & a & a & a  \kill \hline
\[
A
\]
 \par &
\[
  \ \ \gamma
\]
 \par \ \ \ \par  \quad [N/m]&
\[
\omega = \sqrt {\frac{{\gamma} }{{M_{0}} }}
\]
 \par  \quad [{\kern 1pt}s$^{-1}$]&   \par
\[
\hbar {\kern 1pt} {\kern 1pt} \omega
\]
 \ \quad \ \par   [MeV]  \par &
\[
 \  \  \  \lambda \ \ \
\]
  \par \ \ \par \ \ \  \quad [m] \par &
\[
R = R_{0} {\kern 1pt} A^{1/3}
\]
 \par \ \  \par  \quad \ \ [m]&
\[
R/\lambda
\]
 \\
\hline \ \par  \ \,  50 \par \ 100  \par \ 200  \par \ 250
\par & \ \par 16.2 ${\kern 1pt} \times {\kern 1pt}$ 10$^{15}$
\par \ {\kern 1pt} 8.1 $ {\kern 1pt} \times {\kern 1pt}$ 10$^{15}$
\par \  4.05 $\times$ 10$^{15}$
\par \  3.24 $\times$ 10$^{15}$ \par $^{}$& \
\par 3.1 {\kern 2pt} $\times$ 10$^{21}$ \par 2.46 $\times$ 10$^{21}$ \par
1.55 $\times$ 10$^{21}$ \par 1.39 $\times$ 10$^{21}$& \ \par \
2.05
\par \ 1.45 \par \ 1.42 \par \ 0.92 &
\ \par 1.9 {\kern 2pt} $\times$ 10$^{-14}$ \par 2.24 $\times$
10$^{-14}$
\par 2.67 $\times$ 10$^{-14}$ \par 2.83 $\times$ 10$^{-14}$ \par &
\
\par 4.42 $\times$ 10$^{-15}$ \par 5.57 $\times$ 10$^{-15}$ \par
7.01 $\times$  10$^{-15}$ \par 7.56 $\times$ 10$^{-15}$& \ \par \
0.23
\par \ 0.25 \par \ 0.26 \par \ 0.27     \\       \hline

\end{longtable}

\bigskip

In the model considered above the percent composition of protons
and neutrons has not been taken into account and quantities of
protons and neutrons have rather been the same. It is obvious that
the increase in protons should make softer the elasticity of the
inerton field, which, therefore, will decrease to a critical value
$\gamma _{c} $ with further decay of the nucleus. At the same
time, in the case when the quantity of neutrons significantly
prevails that of protons, we cannot expect of the cluster
solution, i.e. a mixture of nucleons will be unable to form weight
nuclei. Of course, taking into account other factors should
correct result (\ref{eq67}). First of all this is the spin
interaction between nucleons, Fermi particles, which should induce
some additional increase in the energy of the system of \textit{N}
nucleons. Nevertheless, the quantitative pattern of the reasons
for the restriction on the weight of nuclei as a function of
\textit{A} has became evident.

\section*{6. Concluding remarks}

\hspace*{\parindent} In the present work, starting from first
principles of the constitution of the real space we have examined
in depth founding physical conditions for the occurrence of
nuclear forces in a system of interacting nucleons. It has been
argued that any canonical particle representing a local
deformation of the space is surrounded by its own deformation
coat. The size of the deformation coat, or crystallite,  is
identified with the Compton wavelength of the particle, $\lambda
_{{\kern 1pt}\rm Com} = h/M{\kern 1pt} c$. And it is this size
$\lambda _{{\kern 1pt}\rm Com}$ that characterizes the effective
radius of nuclear forces. The overlapping of deformation coats of
nucleons results in the induction of some additional deformation
in the space between nucleons, which in an abstract (energy or
phase) space is treated  as the induction of the so-called
potential well.

Due to the interaction of a moving particle with the space
tessellattice, a cloud of spatial excitations called ``inertons"
[12,13] is accompanied the particle. The same takes place in the
case of a moving nucleon when its inertons cloud representing a
substructure of the nucleon's matter waves extends for a distance
$ \sim \Lambda $ (2) from the nucleon. These standing inerton
waves, oscillating with the velocity approaching the velocity of
light $c$, induce a massive (or deformation, or gravitational)
dynamic potential around the particle. In other words, inertons
are quasi-particles that form the relief of the particle massive
potential, which is treated as static. So inertons alone are
responsible for the direct interaction between particles and hence
they are those quasi-particles that ensure short-range action and
reject action at-a-distance forces from quantum systems. The
particle's inerton cloud determines the range (see expression (2))
of application of the wave $\psi$-function formalism employed in
orthodox quantum mechanics. Besides, the availability of the
particle's inerton cloud supports the formalism of hadronic
mechanics developed by Santilli and others [2,3] for strongly
interacting particles: From the submicroscopic viewpoint, the
unification of the kinetic and potential parts of the energy means
the presence of a dynamic field that unifies interacting
particles. It is the inerton field that is capable to ensure such
unification (see, e.g. the structure of the Lagrangian (28)).

Additional characteristics of particles such as the
electromagnetic polarization and spin, which have respective been
clarified in Ref. [14] and Ref. [11], are imposed upon the
particle's structure introducing new properties that manifest
themselves when the particle interacts with its inerton cloud and
other particles.

The above-listed results have allowed us to investigate the major
reasons for the  instability of weight nuclei. The study has shown
that the number of nucleons $A$ that enter into a cluster, i.e.
the nucleus, is defined by condition (67). This condition links
the value of $A$ to the elasticity constant of the inerton field
$\gamma $ in the nucleon, the nuclear density $\rho _{{\kern 1pt}
\rm nucl} $, and the elementary electric charge $e$ of the proton.
The solution (67) depicts that the value of $A$ is inversely
proportional to the elasticity constant $\gamma$ of the inerton
field in a nucleus: an increase in $A$ requires a decrease in
$\gamma$, approaching it to the critical value $\gamma_{\rm c}$
such that at $\gamma < \gamma _{\rm c}$ the inerton field is
incapable of holding nucleons in the cluster, i.e. nucleus, state.

Trying to account for the reasons for nuclear forces, we have
analyzed major views available in the literature including QCD,
quantum field theories, hadronic mechanics, and the Vedic
literature as well. In the epigraph we have quoted the verse from
{\it The \d{R}gveda}\footnote{{\kern 1pt}In modern languages the
word  ``{\it \d{R}gveda}", or ``{\it \d{R}g Veda}", {\kern 1pt}
means the highest knowledge, or the top of knowledge (``{\it
\d{R}g}" is {\it horn} and ``{\it Veda}" is {\it knowledge}).},
which may seem inappropriate, but only at a first glance. Owing to
Roy's decoding [37], it now becomes clear that this is a handbook
on the constitution of the real space, particle physics and
cosmology. In {\it The \d{R}gveda}, the nucleus was encoded under
the name of Sage Vasi\d{s}\d{t}ha, a very respected personage of
ancient Vedic and post-Vedic manuscripts ({\it Vasi\d{s}\d{t}ha}
is translated from Sanskrit as {\it rich, a rich man}). Among
other decodings, we would like to cite the following instances:
God Savit$\bar{\rm a}$, God Varuna, God Mitra, God Aryam$\bar{\rm
a}$ and God Rudra who were interpreted as the
creation-annihilation energy, the electron, the proton, the
neutron and radiation, respectively. Besides, we should note that
the three steps of God Vi\d{s}nu (the universe) have been
explained by Roy as the space web, which consists of 1)
indivisible cells, 2) their interface and 3) the observer space,
i.e. the aggregation of cells. This decoded Vedic pattern exactly
agrees our theory of the real space described above.

Thus the knowledge base of the ancients was inexplicably deep and
we would go ahead keeping in mind their profound views on the
surrounding world. This can help to answer questions that
challenge numerous investigators dealing with nuclear
transmutations that are fixed at low energy reactions today. Sage
Vasi\d{s}\d{t}ha be praised!

\medskip

\subsection*{Acknowledgement}

I am very thankful to an anonymous sponsor who paid the page-charges for this article.

\end{document}